\documentclass[aps,superscriptaddress,showkeys]{revtex4}
\usepackage[utf8x]{inputenc}
\usepackage{amsmath,amsfonts,amssymb,graphicx,color,times,psfrag}
\usepackage{bbold}
\usepackage{epsf}
\usepackage{mathptmx}
\usepackage{subfigure}
\usepackage{dcolumn}
\usepackage{bm}
\usepackage{latexsym}
\usepackage[english]{babel}
\usepackage{epsf}
\usepackage{natbib}
\usepackage{epstopdf}
\usepackage{appendix}
\usepackage[colorlinks=true,citecolor=blue,linkcolor=blue]{hyperref}
\usepackage{ulem}

\newcommand{\bra}[1] {\langle{#1}\vert}
\newcommand{\braket}[2] {\langle{#1}\vert{#2}\rangle}
\newcommand{\ket}[1] {\vert{#1}\rangle}

\global\long\def\proj#1{\mbox{\ensuremath{|#1\rangle\!\langle#1|}}}

\begin{document}


\title{Optical Abelian Lattice Gauge Theories}
\author{L. Tagliacozzo}
\email{luca.tagliacozzo@icfo.es}
\affiliation{ICFO  The Institute of Photonic Sciences Av. Carl Friedrich Gauss, num. 3, E-08860 Castelldefels (Barcelona), Spain}

\author{A. Celi}
\email{alessio.celi@icfo.es}
\affiliation{ICFO  The Institute of Photonic Sciences Av. Carl Friedrich Gauss, num. 3, E-08860 Castelldefels (Barcelona), Spain}

\author{A. Zamora}
\affiliation{ICFO The Institute of Photonic Sciences Av. Carl Friedrich Gauss, num. 3, E-08860 Castelldefels (Barcelona), Spain}

\author{M. Lewenstein}
\affiliation{ICFO  The Institute of Photonic Sciences Av. Carl Friedrich Gauss, num. 3, E-08860 Castelldefels (Barcelona), Spain}
\affiliation{ICREA-Instituci\'{o} Catalana de Recerca i Estudis Avancats, 08010 Barcelona, Spain}

\begin{abstract}
We discuss a general framework for the realization of a family of Abelian lattice gauge theories, i.e., link models or gauge magnets, in optical lattices. 
We analyze the properties of these models that make them suitable for quantum simulations. Within this class, we study in detail the phases of a U$(1)$-invariant lattice gauge theory in 2+1 dimensions, originally proposed by P. Orland.  By using exact diagonalization, we extract the low-energy states for small lattices, up to $4\times 4$. We confirm that the model has two phases, with the confined entangled one characterized by strings wrapping around the whole lattice. We explain how to study larger lattices by using either tensor network techniques or digital quantum simulations with Rydberg atoms loaded in optical lattices, where we discuss in detail a protocol for the preparation of  the ground-state. We propose two key experimental tests that can be used as smoking gun of the proper implementation of a gauge theory in optical lattices. These tests consist in verifying  the absence of spontaneous (gauge) symmetry breaking of the ground-state and the presence of charge confinement. We also comment on the relation between 
standard compact U$(1)$ lattice gauge theory and the model considered in this paper.

\end{abstract}

\maketitle

\section{Introduction}

Studying the strongly correlated regime of  quantum many-body systems (QMBS) remains  a very challenging task. We do believe that most of the  interesting physics associated to them  can be captured  by simple models described through Hamiltonians involving  interactions among only few of their constituents. 
However, even for such simple models, we are not capable of deriving their properties, as we still miss computational tools that are generically applicable.

In few fortunate cases, the models constructed are exactly solvable, and almost all physical quantities can be computed exactly. When this is not the case, one needs to appeal to numerical simulations.  A major obstacle to this approach  is provided by the fact that the complexity of simulations increases exponentially with the number of the QMBS  constituents. This means that, typically, one can only deal exactly with few (say up to forty) constituents.

Some larger QMBS can still be simulated approximately. This occurs, for instance, in weakly interacting systems, where  perturbation theory (spin-wave and its modifications), mean-field approaches and density functional theory are particularly successful. In all these cases, indeed, there is a  small parameter -- namely the interaction among constituents--  that can at the beginning be neglected and then systematically included in the model \cite{fiolhais_primer_2003,martin_electronic_2004}.

The strongly interacting regime of QMBS is much more complex to address for the lack of such small parameter. However, the equilibrium properties of bosonic non-frustrated systems can be extracted using Quantum Monte-Carlo (QMC) techniques \cite{Sorella-becca,sandvik_computational_2011}.
Frustrated and/or fermionic systems are more difficult to study, but  great effort has also been dedicated to them. At present, however, different methods often provide different results so that physical properties of these models are  in most cases a subject of controversy (see i.e. \cite{Sorella-becca} vs. \cite{corboz}).

For weakly  entangled states, tensor networks algorithms (TN) are emerging as an alternative to other techniques such as QMC, and in some cases can also be merged with the latter  to provide new numerical tools \cite{stoudenmire_one-dimensional_2011,sandvik_variational_2007,ferris_perfect_2012,schuch_simulation_2008,wang_monte_2011}.
TN aim at providing a variational ansatz to describe low-energy states of QMBS, with resources that scale at most polynomially with the number of constituents. 
These techniques also give access to short-time dynamics of QMBS \cite{vidal_efficient_2003,jeckelmann_dynamical_2002}, but for the moment still fail in describing their long-time, out of equilibrium, dynamics.

In this situation, analog quantum simulations (AQS) appear as an alternative tool to study QMBS \cite{trotzky_probing_2011,cheneau_light-cone-like_2012}. These are experimental setups with quantum constituents (photons, ions, atoms) that are used to encode the desired QMBS \cite{maciej_book}. State of the art control of these quantum systems allows also to artificially induce the desired many-body Hamiltonian acting  on the constituents.

Gauge theories are special types  of QMBS, where states of the system are forced to obey local symmetry constraints.
They stand at the core of our understanding of known fundamental interactions (electro-weak, strong, and, in a broader sense, gravitational).
They are also candidates to describe  the low-energy physics of frustrated antiferromagnets \cite{misguich_quantum_2008,balents_spin_2010}. 
As for other QMBS, their strong coupling regime is not completely understood: in particular, they present puzzling phases, such as, confined phases and topological phases, which escape the standard Landau paradigm of phase classification.

For gauge theories, and their version defined on (spatial) lattices (LGT),  in the absence of fermions --  or when the number of fermions is such that there is no sign problem--   one can perform Monte-Carlo simulations. Recently, ideas about how to perform tensor network simulations of such theories, have been push forward  as well \cite{tagliacozzo_entanglement_2011}.

Applying   AQS to study  gauge theories is particularly challenging since, in general, their dynamics involve Hamiltonian with couplings among more than the nearest-neighbors, and more than just two-body terms.  This is the reason why, at present, the community has mainly focused on simulating the effects of static non-trivial background gauge fields on the phases of the matter. In this limit, the many-body interactions can be neglected, and the effect of the gauge field reduces to a specific modifications of the kinetic term of the matter field -- hopping in the lattice--  that only involves two-body interactions.

In this context, most of the  studies deal  with  ultracold atoms \cite{maciej_book}, and some of the theoretical proposals have been already realized in the experiments on the lattice \cite{Aidelsburger2011}, or, off the lattice \cite{lin2009,dalibard2011}. In optical lattice simulations,  one can  tailor the hopping phase/matrices properly, mainly by optical means (but also by lattice time-modulation  or  rotation \cite{struck2011, bloch2008}). The interest in such studies ranges from Quantum Hall physics, where strong  magnetic fluxes  -- very hard to obtain in condensed matter with real magnetic field--  are needed \cite{tsui1982,laughlin1983}, to the simulation  of relativistic matter and topological insulators, or extra-dimensions \cite{Mazza2011,boada_quantum_2012}.


A natural step further towards the full simulation of gauge theories with  AQS,  is  to include the dynamics of the gauge fields.  Such aim, for instance in two dimensions, requires the engineering of  at least three-body couplings in triangular lattices and four-body couplings in the more common square lattices. These kind of interactions are very difficult to induce in AQS, such as,  ultracold optical lattices \cite{buchler_three-body_2007}). 
There have already been some proposals to overcome these difficulties for Abelian gauge fields using molecular states \cite{buchler2005}, and, more recently, BECs and single atoms with internal states in \cite{zohar2011} and \cite{zohar_simulating_2012}, respectively. 

Here, we pursue an alternative solution. We propose to simulate  the dynamics of gauge fields by implementing digital quantum simulations with Rydberg atoms (DQRS).  Rydberg atoms are neutral atoms that can be excited to states close to the continuum spectrum, having very strong dipolar moments. Such dipoles induce long-range interactions that make possible the simultaneous interaction of several atoms together. 
The basic ideas for DQRS are described in  \cite{Muller09} and involve the presence of ancilla atoms, apart from  the ones entering the Hamiltonian of the  system to be simulated. These are called ``control'' Rydberg atoms. The other atoms, whose interaction should encode the Hamiltonian to be simulated, are generally called  ``ensemble'' Rydberg atoms. They have to be physically arranged following the pattern of the  many-body interactions appearing in the desired Hamiltonian. For example, if one is trying to encode four-body interaction of the four atoms around an elementary plaquette of a square lattice,  such  atoms  should be   inside the  blockade radius of a given control atom, whose dipole moment is   used to implement the wanted plaquette interaction among them.

One can carry out simultaneous operations on all the atoms inside the blockade radius of a control atom,  operations determined by the state of the control atom itself.  This can be achieved through  a laser setup involving two- and three-photons transitions.
In particular, one can engineer 2D lattices where the ensemble atoms are arranged on the links, while the control atoms are at each site and  at the center of each plaquette; at same time, one can tune the lattice spacing so that all the links belonging to a plaquette and entering a site  are simultaneously contained inside the blockade radius of the respective plaquette and site control qubits. With this geometry one can  perform  LGT simulations \cite{Weimer10}.

In this setup, one can perform arbitrary time-evolution with the desired (Abelian) LGT Hamiltonian, and,
in some simple cases, one can also accomplish dissipative quantum simulations to prepare a desired state. This is, for instance, the technique proposed for the preparation of  the ground-state of the Toric Code in  \cite{Weimer10}, whose low-energy physics,
 in the appropriate limit,  can be mapped to the  $\mathbb{Z}_2$ LGT. 

In this paper, we provide  a specific proposal to perform a DQRS of a U$(1)$ LGT in two dimensions.  An important ingredient of the proposal  is the choice of a specific  U$(1)$ LGT where the constituents are qubits. Such qubits can be represented in terms of atoms with two logical level states, such as the ones available in standard Rydberg setups. The corresponding  formulation of U$(1)$ LGT is  also known as link model or gauge magnet \cite{horn_finite_1981,orland_lattice_1990,Chandrasekharan:1996ih,Brower:1997ha}.

The outcomes of our proposal are multifold. 
First, the alternative formulation that we consider is not only more practical to be implemented -- the local Hilbert space is finite--  but also -- at least it is our belief--  closer to the language  of the QMBS community than the original Hamiltonian formalism proposed by Kogut and Susskind \cite{kogut_hamiltonian_1975}.  
Within this general framework, we show how to derive the specific $U(1)$ gauge magnet we are interested in. 

Furthemore, we introduce an exactly solvable version of gauge magnet, where we are able to provide the exact ground-state wave function. We use this example to illustrate two paradigmatic properties of gauge theories, i.e., the absence of spontaneous breaking of the gauge symmetry, and the presence of charge confinement.  The quantitative check that such properties hold can be used by experimentalists as a {\it sine qua non} test of the quality of the simulator. 

Finally, we determine a specific protocol to prepare the ground-state of the model and perform arbitrary, out of equilibrium, dynamical simulations.
As well known, one can extract the excitation spectrum of the model by performing these simulations.

Our approach is relevant  because it allows the simulation of a LGT with continuous symmetry groups away from the special points, as studied in  \cite{Weimer10}.  It provides a generalization of the techniques described in  \cite{Verstraete09,diehl_quantum_2008,kraus_preparation_2008}, such to permit to prepare ground-states of Hamiltonians that are not frustration-free.
Furthermore, we  explain how to perform out of equilibrium long-time-evolution of the U$(1)$ gauge magnet with a DQRS, following the prescription of  \cite{Weimer10}.

It is worth to notice that our ground-state preparation scheme successfully merges several available techniques. It is, indeed, made of two steps. First, by taking advantage of the fact that the square lattice is bipartite, we decompose the Hamiltonian of the gauge magnet as the sum of two Hamiltonians, each of them for separated is frustration-free. This allows us to prepare the ground-state any of the two with standard dissipative protocols. 
The  second step involves  a driven evolution via DQRS to the final Hamiltonian of gauge magnets. The turning  on of the part of the Hamiltonian that   has been neglected in the first instance,  can be adiabatic or not. Indeed, the efficency of the ground-state driving can be improved by following the chopped random
basis (CRAB) protocol \cite{caneva_chopped_2011,doria_optimal_2011}, which allows for closed-loop optimization experiment, i.e., the preparation time is optimized recursively by running the DQRS. 
In this way, we manage to prepare the ground-state of the complete Hamiltonian, a very complex, strongly interacting, non frustration-free, i.e., frustrated Hamiltonian. In practise, such preparation can be applied to any system governed by a local Hamiltonian in bipartite lattices.

The paper is organized as follows. In section \ref{sect:constr}, we introduce the concepts necessary to formulate a LGT. In doing so, we assume that the reader is familiar with many-body quantum systems, while the knowledge of continuous or Lagrangian formulation of gauge theory is not assumed. In particular, we perform a constructive derivation of the Hilbert space and Hamiltonian of an arbitrary Abelian LGT (this section is complemented by the examples contained in  the Appendix). In section \ref{sect:magn}, we review the properties of the  U$(1)$ gauge magnets as a specific instance of generic Abelian LGT using the language and the notation  introduced in the previous section. We discuss how to split the Hamiltonian in the sum of two frustration-free Hamiltonians.  This is a crucial step both for applying the QDRS protocol of the ground-state preparation, and for analyzing the absence of spontaneous symmetry breaking and the presence of charge confinement.  We propose  these phenomena as key experimental observables to probe whether the simulator is working properly.
In section \ref{sect:ryd}, we start by reviewing the ideas beyond digital quantum simulation and the mesoscopic Rydberg gate \cite{Muller09}. We discuss in detail the setup necessary to perform a DQRS of the U$(1)$ gauge magnet time-evolution, which permits the study of long-time, out of equilibrium, dynamics. We conclude by giving  special emphasis on how to prepare the ground-state of the U$(1)$ gauge magnet with a mixture of dissipative and adiabatic evolution and discuss the relation between our model and standard compact U$(1)$ LGT. We propose two possible sets of measurements that can be used by the experimentalists to check both absence of spontaneous symmetry breaking and charge confinement. For reader's convenience, the main features of  the experimental set-up and the experimental sequence are summarized in section \ref{summary}.

\section{Constructing an Abelian lattice gauge theory}\label{sect:constr}
We introduce an Hamiltonian  formulation of Abelian lattice gauge theories. 
Gauge theories are generally  studied in the Lagrangian formulation, by introducing an action invariant under local transformations, and quantizing it via the path integral formulation of Quantum Mechanics \cite{montvay_quantum_????, creutz_quarks_1985}. By discretizing the space-time, one obtains a lattice description  that can be treated with the tools of statistical field theories. 
In this formulation, the corresponding quantum Hamiltonian  can be derived from the transfer matrix of the model, as as originally proposed by  Kogut and Susskind \cite{kogut_hamiltonian_1975} (in the same way one can map the classical 2D Ising model at finite temperature on to the quantum Ising chain at  the zero temperature with a transverse field \cite{kogut_introduction_1979}).
 
Here, we take an alternative route. Indeed, an Abelian lattice gauge theory  is a particular case of a many-body quantum system, where the constituents are arranged on a lattice and the states (and the observables) are invariant (as we will explain below) under local transformations, which are elements of a given  Abelian group $G$.  Starting by this operative definition, we can construct an Hamiltonian for a LGT. 
It is worth to notice that, for a given $G$, infinitely many gauge invariant theories -- and Hamiltonian--  can be defined, although many of them may be equivalent under renormalization group flow, i.e., may share the same fixed point and critical properties. 
  Baring this consideration in mind, the 
  basics steps to follow to construct a gauge invariant Hamiltonian for a lattice system, are: i) the choice of the gauge group $G$ that, as we will see,  has immediate consequences on the  definition of  the Hilbert space of the model; ii) the definition of the operators defining the gauge transformations; and iii) the determination of the  set of operators that can be used to build the Hamiltonian, which  has to be  invariant under gauge transformation.
This constructive approach  should sound more  familiar   to people working in the field of quantum many-body systems and condensed matter rather than the one originally proposed by Kogut and Susskind, as it is not requiring a prior knowledge of  gauge theories. It also makes explicit that each of the points i)-iii) entails a large freedom. The original  Kogut-Susskind proposal reveals itself as a specific choice of LGT. Also, using this approach one can interpret  different  LGT  known in the literature as link models or gauge magnets  \cite{horn_finite_1981,orland_lattice_1990,Chandrasekharan:1996ih,Brower:1997ha} as a truncated version of the original Kogut-Susskind proposal.
We would like to mention that this approach is rooted  in the celebrated works by Kitaev about the  Toric Code and Quantum Doubles \cite{kitaev_fault-tolerant_1997} that we  follow closely. 
When defining the steps i)-iii)  we will explicitly construct  the simplest $\mathbb{Z}_2$ LGT, so that the reader will have an immediate example on which she/he can understand the more general abstract formulation. Several other examples are contained in the Appendix.

\subsection{Gauge invariance, the gauge invariant Hilbert space and operators}
\emph{i) The Hilbert space.}
In order to construct a LGT one first need to specify the symmetry group  $G$. For example, the simplest choice is to consider  $G=\mathbb{Z}_2$, which is the group formed by two elements, $1 ,e$, with the following multiplication table $1\times e = e\times 1=e, e\times e=1\times 1= 1$.

The constituents of the LGT are attached to the links $l$ connecting sites $s$ of an oriented lattice $\mathcal{L}$. 
The fact that the lattice is oriented  means that, when addressing a link, one has not only to specify its position but also its orientation (the direction to follow in order to walk along it, i.e., in 2D  either from left to right or from up to down or viceversa). The need for considering oriented lattices can be understood through the physical picture of the links as currents moving from a site to another, the orientation is the direction of the current (and the gauge condition is the analogue of the conservation of the current at each site).  

Each of the constituents attached to the oriented links is described by a vector in a local Hilbert space $\mathcal{V}_{l}$.
If the lattice has $N$  links, the global Hilbert space is defined as the tensor product ${\cal H} = \mathcal{V}_{l} ^{\otimes N}$. 

The choice of $\mathcal{V}_{l}$ is not unique, and this is the  first freedom at hand in the construction of LGT.   
In particular, the original formulation by Kogut and Susskind uses as local Hilbert space the group algebra $\mathbb{C}(G)$, so that the space of constituents is the vector space generated by linear combinations  of elements of $G$ with complex coefficients. To each element of the group $G$, one associates a vector $\ket{g}$, with the property $\braket{h}{g}=\delta_{g,h}$. Here $\delta_{.,.}$ is the Kronecker delta. In the simplest case of $\mathbb{Z}_2$, the local Hilbert space is generated by the two orthogonal vectors $\ket{1}$ and $\ket{e}$, and it is isomorphic to $\mathbb{C}^2$. 
  In the Kogut-Susskind formulation, the dimension of the local Hilbert space is strictly related to the number of elements in the group. Hence, it becomes infinite for continuous groups like $U(1)$. However, as we will show below, it is perfectly consistent to consider a local Hilbert space whose dimension does not depend on the number of elements $G$.

\begin{figure}
 \includegraphics[height=5cm]{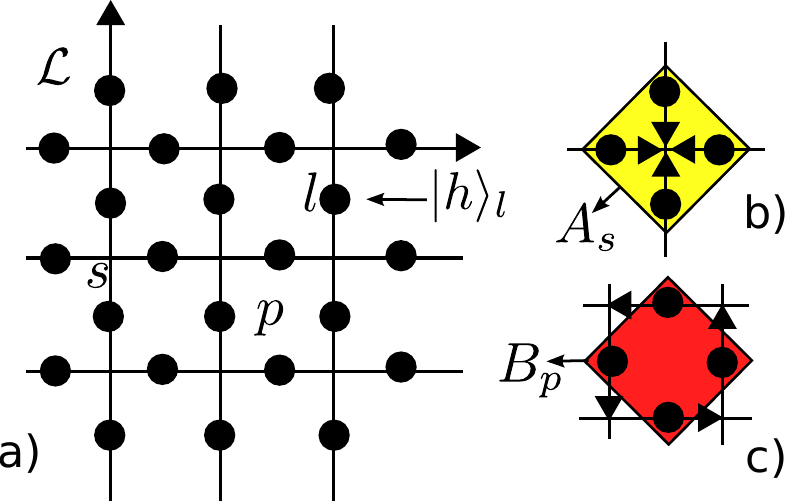}
\caption{  a) A lattice gauge theory is many-body system defined on a lattice  $\mathcal{L}$. The constituents $\ket{h}_l$ are attached to the links $l$. The lattice is oriented, meaning that links inherit the standard orientation of the embedding space, depicted in the figure as arrows. When defining operators acting on the states attached to links, one has also to specify an orientation. b) and c) The building blocks of a LGT include There  two sets of four-body operators. b) The one attached to sites $s$, $A_s$, which we also refer to as \emph{star} operators,  are  used to impose the gauge invariance. c) The ones attached to plaquettes $p$ $B_p$, which we also refer to as \emph{plaquette} operators,  describe the dynamics. We define both $A_s$ and $B_p$ by using a customary oriented star and plaquette. \label{fig:def} }
\end{figure}

As an example, the space of constituents  of a 2D  lattice is  sketched in Fig. \ref{fig:def}.
Since we are dealing with an oriented lattice, we also orient the states. Following physical intuition of a links as currents between sites, the change of orientation corresponds to {\it an inversion} of the currents.  
With the standard choice of the local Hilbert space as group algebra, $\mathcal{V}_{l}$ = $\mathbb{C}(G)$, the physical intuition translates in associating the inverse in $G$ to a flip of the orientation. This can always done since the states are elements of a group that by definition have their own inverse. Hence, changing the orientation of a link is equivalent to the operation
\begin{equation}
\ket h_{l}\to\ket{h^{-1}}_{-l}. \label{or}
\end{equation}
The detailed graphical explanation of the role of the orientation is given in the Appendix \ref{sect:z3}, Fig. \ref{fig:Def} iv). 
It is worth to observe that, for the guiding example of  $G=\mathbb{Z}_2$, the orientation plays no role since $e^{-1}=e$ and $1^{-1}=1$, hence,  $\ket e_{l}=\ket{e}_{-l}$ and $\ket 1_{l}=\ket{1}_{-l}$.

The only requirement on the local Hilbert space, $\mathcal{V}_{l}$, needed to construct a consistent LGT, is that $\mathcal{V}_{l}$ has to be isomorphic to a representation of the symmetry group $G$. In particular, the choice of  $\mathbb{C}(G)$  induces the choice of  the fundamental representation of the group, $R_f(G)$. The action of an element of $X(g) \in R_f(G)$ on  $\ket{ h}$, is defined as
\begin{equation}
X(g)\ket {h} =\ket {(g h)},
\end{equation}
This means that  the matrix representation of  $X(g)$ reads
\begin{equation}
X(g) =\sum_{h \in G}\ket {g h}\bra{h}.
\end{equation}
For $G=\mathbb{Z}_2$, it follows that $X(1)=\proj{1}+\proj{e}$ and $X(e)=\ket{e}\bra{1}+\ket{1}\bra{e}$, which coincide with the identity matrix and the $\sigma^x$ Pauli matrix, respectively. 

When  we change the orientation of a link, accordingly to  (\ref{or}),  we also have to invert the operator that is acting on it,
\begin{equation}
\ket {(g h)}_l \to \ket {(h^{-1}g^{-1})} _{-l} =  X_{-l}(h^{-1}g^{-1} h) \ket{h^{-1}}_{-l} =  X_{-l}(g)^{\dagger} \ket{h^{-1}}_{-l}\label{or_ope}
\end{equation}
where $l$ is an arbitrary oriented link of ${\cal L}$ and  we have used the fact that, for Abelian groups, $h^{-1}g^{-1} h = g^{-1}$, and,  for unitary representations,  $X(g^{-1})=X(g)^{\dagger}$. The irrelevance of  the orientation in the $\mathbb{Z}_2$ case translates in that both $X(1)$ and $X(e)$ are Hermitian.

\emph{ ii) Gauge invariance and the physical Hilbert space.} 
A generic gauge transformation is defined by choosing a group element for each site of the lattice and rotating all the links entering that particular site with the matrices representing the rotation for that particular element. The building blocks of gauge transformations are thus obtained by acting on all  links entering a given site with $X(g)$.
The operator that induces such rotation is often referred as \emph{star} operator and for the specific case of a 2D square lattice is defined as
 \begin{equation}
A_{s}(g)\equiv \bigotimes_{l\in \{l_i\}_s}X_l(g):\,\ket{h_{l_{1}},h_{l_{2}},h_{l_{3}},h_{l_{4}}}\to\ket{gh_{l_{1}},gh_{l_{2}},gh_{l_{3}},gh_{l_{4}}},\end{equation}
where $s$ is a generic site  of the lattice and   $\{l_i\}$, $i=1,\dots,4$, is the set of all the links {\it entering} $s$, as depicted on  Fig. \ref{fig:def} b). In view of  (\ref{or}) and  (\ref{or_ope}),  this operator acting on the links of a lattice oriented in  the  conventional way (such as in Fig. \ref{fig:def}a) becomes 
\begin{equation}
A_{s}(g):\ket{h_{-l_{1}},h_{-l_{2}},h_{l_{3}},h_{l_{4}}}\to\ket{h_{l_{1}}^{-1}g^{-1},h_{l_{2}}^{-1}g^{-1},gh_{l_{3}},gh_{l_{4}}},
\end{equation}
where the links are numbered clockwise starting by the one on the top. In terms of $X(g)$ operators, the above expression is equivalent to  $A_s(g)=X^\dagger_{l_1}(g)X^\dagger_{l_2}(g)X_{l_3}(g)X_{l_4}(g)$. The choice of the group element $g$ can vary from site to site so it would be more appropriate to use $g_s$ but in order to simplify the notation we stick to $g$. 
For $G=\mathbb{Z}_2$, the only non-trivial $A_s$ operators are those associated with the group element $e$ and are defined as $A_s(e)=\sigma^x_{l_1}\sigma^x_{l_2}\sigma^x_{l_3}\sigma^x_{l_4}$. The action of $A_s(e)$ then induces  spin-flips on all qubits entering the specific link $s$.

States are  invariant under the gauge transformations at site $s$ if they are eigenvectors of the $A_{s}(g)$ operators with eigenvalue $+1$,
\begin{equation}
 A_{s}(g) \ket{\psi} =\ket{\psi}, \forall g \in G. \label{eq:gauge_cond}
\end{equation}
Since generic gauge transformations are product of local ones, a state is  gauge invariant if 
\begin{equation}
{\cal T}(\{g\})\ket{\psi} \equiv \bigotimes_{s\in\mathcal{L}}A_{s}(g)\ket{\psi} = \ket{\psi},  \forall g \in G.
\end{equation} 
This leads to the definition of the \emph{physical} Hilbert space, ${\cal H}_G$, as the set of those states in {\cal H} that are gauge invariant,
\begin{equation}
{\cal H}_G=\{\ket \psi\}\ s.t.\ {\cal T}(\{g\}) \ket{\psi}=\ket{\psi},\, \forall g\in G. \label{eq:gauge_invariant_state}
\end{equation}
It is important to notice that the \emph{physical} Hilbert space is a subspace of the original Hilbert space,
\begin{equation}
{\cal H}_G\in {\cal H}\equiv\mathcal{V}^{\otimes N}\simeq{\cal C}^{|G|\otimes N},
\end{equation}
For $G=\mathbb{Z}_2$, the requirement of invariance at a specific site under the action of $A_s(e)$ reduces the dimension of the Hilbert space from $16$ to $8$. In order to determine the allowed states, it is better to diagonalize $\sigma^x$, $\sigma^x=\proj{+}-\proj{-}$. In this basis, all those states formed as tensor product of four eigenstates of $\sigma^x$ with an even number of $\ket{-}$ are eigenstates of $A_s(e)$ with eigenvalue $+1$, and thus gauge-invariant. As we will describe in the following, one can give a nice geometrical interpretation to those states in terms of closed string of $\ket{-}$. 

\emph{ iii) Operators compatible with the requirement of gauge invariance.}
Once defined the \emph{physical} Hilbert space, we focus on the operators that are compatible with the requirement of gauge invariance. By definition, the LGT Hamiltonian $H$ has to respect the local symmetry, i.e., commute with all  \emph{star} operators  $A_{s}(g_{s})$,
\begin{equation}
[H, A_s] = 0, \ \forall A_s. \label{eq:comm_ham}
\end{equation}
Our goal is to built up $H$ as a sum of local terms.

To this aim,  we introduce a set of operators  $Z(r)$,  acting on the tensor product of $\mathcal{V}_{l} \otimes \mathcal{V}_g$. Again, there is a lot of freedom in defining  the factor  $\mathcal{V}_g$, since the only requirement is to support an irreducible representation (irrep) of the gauge group (for a definition of irreducible representation refer to \cite{tinkham_group_2003}). When the local Hilbert space is the group algebra, the operators $Z(r)$ acquire the following form   
\begin{equation}
Z(r)  =\sum_{h\in G} R_r(h) \otimes \proj {h}, \label{eq:def_z}
\end{equation}
where $R_r(h)$ is the matrix representing $h$ in the irreducible representation $r$ of $G$. Since all irreducible representation of Abelian groups  are one dimensional  and isomorphic, for Abelian theories we can drop the index $r$ from  $R_r$ and  think of them as acting only on  $\mathcal{V}_{l}$ since, in this case, $R(h)$ is just a phase.  For example, in the case of $G=\mathbb{Z}_2$, the only non-trivial choice for $Z$ is $Z=\sigma^z=\proj{1}-\proj{e}$.

The relation between  $Z$ and  $X(g)$ is encoded in the commutation relation
\begin{equation}
Z X(g)\ket {h} =R(g h) \ket {g h} = R(g) X Z \ket {h} \label{eq:comm_z_x}
 \end{equation}
This immediately suggests  a   minimal choice to fulfill (\ref{eq:comm_ham}). We can, indeed, consider as  building blocks for  $H$  the product of four $Z$ operators acting on links around elementary  plaquettes of the lattice  
\begin{equation}
B_p\equiv \bigotimes_{l\in\{l_i\}_p} Z_l(r),\label{Bp}
\end{equation} 
that we generally call \emph{plaquette operator}.
Here, $\{l_i\}_p$, $i=1,\dots,4$, is a set of links belonging to the plaquette $p$ and anti-clockwise oriented starting from bottom, as sketched in Fig. \ref{fig:def}c.  By rewriting $B_p$ for the standard-oriented lattice, as in Fig. \ref{fig:def}a, we obtain
\begin{equation}
B_p=Z_{l_1}(r)Z_{l_2}(r)Z(r)^\dagger_{l_3}Z(r)^\dagger_{l_4}. 
\end{equation}
In the elementary case of $G=\mathbb{Z}_2$, the plaquette operator is just $B_p=\sigma^z_{l_1}\sigma^z_{l_2}\sigma^z_{l_3}\sigma^z_{l_4}$.

As a plaquette and a star operator share none or two links, $l$ and $l'$, it is sufficient to verify the relation 
\begin{equation}
 [Z_l(r) \otimes Z_{l'}(r), X_l(g) \otimes X^{\dagger}_{l'}(g)]=0, 
\end{equation}
in order to check  that $B_p$ and $A_s$ commute. As the above relation is a direct consequence of (\ref{eq:comm_z_x}) (and that for any representation $R(g^{-1})=R^{-1}(g)$), the desired result
\begin{equation}
 [B_p, A_s(g)] =0 , \quad \forall \{p, s\} \in \mathcal{L}, g \in G,
\end{equation}
 holds.

As obvious for an Abelian gauge group -- the case of interest in this work--  any operator $X_l(h)$ commutes with $A_{s}(g_{s})$. Hence, it follows that any hermitian functional of $B_p$ and $X_l(h)$ is a good gauge invariant Hamiltonian for Abelian gauge theories. In particular, we focus on the linear combination
\begin{equation}
H(\theta)= - \cos\theta  \sum_p B_p + \sin\theta \sum_l  X_l(g) + H.c.\,\label{eq:ham_ks},
\end{equation}
where $p$ are the elementary plaquettes of the lattice and $l$ are the links. This coincides with the Kogut-Susskind Hamiltonians \cite{kogut_hamiltonian_1975}. In the specific case of $G=\mathbb{Z}_2$, it can be written explicitly in terms of Pauli matrices
$H(\theta)= - \cos\theta  \sum_p \prod_{l\in p} \sigma^z_l + \sin\theta \sum_l  \sigma^x_l(g)$.

On physical ground, the $B_p$ and $X_l$ operators describe magnetic and electric interactions, respectively. 
For the specific example of $G=\mathbb{Z}_2$, both electric and magnetic fluxes can only have two values, $\pm 1$.

It is worth noticing that the presence in the Hamiltonian of terms involving  four-body interactions is a direct consequence of gauge invariance. There have been several attempts to obtain a gauge invariant Hamiltonian starting from a model with only first neighbour interactions. For instance, by defining two-body Hamiltonians on a coarse grained lattice \cite{ocko_nonperturbative_2011}, one can obtain in some case equivalent Hamiltonians to those with four-body interactions. However, at the moment these type of constructions are restricted to exactly solvable models, and  it is unclear how to generalize them to a generic   Hamiltonian, as the one we will consider in the following. Thus, in this paper we stick to the idea that gauge invariance (on square lattices) requires four-body interactions.

\subsection{Gauge magnets or link models}\label{sect:magn}

Up to this point, we have seen the general requirements for obtaining a LGT. That is to say, a LGT is characterized by  i) the presence of a gauge symmetry that permits to ii) identify a subspace of the full Hilbert space as the \emph{physical} Hilbert space where iii) only gauge invariant operators can be considered.  In order to get the original Kogut-Susskind proposal, we need to make specific choices for all the three points. Indeed, for i) we use as a local  Hilbert space the group algebra $\mathbb{C}(G)$. This immediately implies that in  ii) the $X(g)$ operators  are  obtained by considering the regular representation of the rotation by an element $g$ of  group $G$. Furthermore, in iii) we have constructed the Hamiltonian using the simplest closed path of the lattice, the plaquettes. Nevertheless, any of the above choices may be changed, resulting in a different gauge theory. Here,  we focus  on alternative choices  to i).  The  reason is that the $\mathbb{C}(G)$ algebra for a continuous 
group is infinite dimensional, while the implementation we have in mind of LGT simulator with optical lattices can only deal with finite dimensional local Hilbert spaces (see section \ref{sect:ryd}).

Our aim is to construct an Abelian LGT with the smallest possible local Hilbert space, independently on the number of elements of the group $G$. This  leads to LGT that have been called gauge magnets or link models \cite{horn_finite_1981,orland_lattice_1990,
Chandrasekharan:1996ih,Brower:1997ha} in the literature, whose particular case is that $U(1)$  LGT with finite dimensional local Hilbert space. 

Let us start by comparing those LGT to the formulation of Kogut-Susskind. In order to do so, let us consider again  the (possibly infinite dimensional) local Hilbert space    $\mathbb{C}(G)$.
A celebrated theorem   of group theory \cite{tinkham_group_2003}, allows us to truncate  $\mathbb{C}(G)$ to a finite dimensional Hilbert space,  without destroying the gauge symmetric structure. The theorem states that the regular representation (the one that acts on  $\mathbb{C}(G)$) can be written as the direct sum  of all possible irreducible representations (with a multiplicity equal to their dimensionality). 

As a consequence of this theorem, we are guaranteed that there is a   change of basis such that the operators $X(g)$  become block diagonal,
\begin{equation}
  X(g) = \oplus_r\sum_{ i j} R_r(g)^{i,j} \ket{(r, i)}\bra{(r, j)},     
\end{equation}
where we have explicitly written the block structure of $X(g)$ labeled by the  $r$  irrep, and   the  $R_r(g)^{i,j}$ are matrix elements of the irrep $r$ of $g\in G$, $i,j=1 \cdots dim(r)$. We call the rotation matrix that brings all  $X(g)$ to the block diagonal  $\alpha((r, i),g) $,  so that we can express the  $Z(r)$  operators in the new basis as  
\begin{equation}
 Z(r) = \sum_{(p,i), (q,j)} z(r)^{(pi),(qj)}\ket{(p,i)}\bra{(q,j)}.
\end{equation}
where the matrix $z(r) ^{(pi),(qj)}$ is $z(r) ^{(pi),(qj)}=\sum_g \alpha((p,i),g) R_r(g) \alpha^{-1}(g,(q,j))$. We call the new basis with abuse of notation $\{\ket{r}\}$ basis. 

In the  $\{\ket{r}\}$ basis, we can safely truncate the local Hilbert space of the LGT without any effect on the symmetry requirements. Indeed, we just have to include at least one of the diagonal blocks of $X(g)$. However, it could happen that by keeping just one block either $X(g)$ or $Z(r)$ become trivial (that is indeed the case for Abelian theories). For this reason, we need to keep at least two irreps. In this case, the local Hilbert space will have dimension $d= dim(r_1) + \cdots dim(r_n)$, where one can stop at the first $n$ that  provides both non-trivial symmetry requirement, and non-trivial dynamics.

In the following, we will provide the specific example of the $U(1)$ gauge magnets, since it is the one we are interested in simulating with optical lattices.  In  the Appendices, we provide further examples for  generic   $\mathbb{Z}_N$ gauge magnets. 

\subsection{U$(1)$ Gauge magnet}\label{truu1}

Using the ideas of the previous section, we can easily study the U$(1)$ gauge magnet. A gauge magnet differs from the standard LGT by the choice of the local Hilbert space. Rather than using the group algebra, one retains  at least two arbitrary irreps, $r_1\oplus r_2$.  Since U$(1)$ is Abelian, its irreps  are one-dimensional: it means that  we can work  with qubits. We could choose any irreps $r_1$ and $r_2$, the simplest choice is to select the trivial irrep where all the element are mapped to $1$, and the irrep where a given element $g\in$U$(1)$, distinct than the identity $\mathbb{1}$,  is mapped to the phase $e^{i\alpha_g}$. As we are interested in faithful irreps, $\alpha_g$ has to be chosen not commensurable with $2\pi$, i.e, $\frac{\alpha_g}{\pi}\notin \{$Rationals$\}$, if $g$ is s.t. $g^n\neq \mathbb{1}$, $\forall n\in \mathbb{N}$. Note that asking local invariant under a $g$ as such it is sufficient to ensure invariance under any local U$(1)$-transformation. We denote the eigenvectors of the $X(g)$s as $\ket +$ and $\ket -$.  This means that the $X$ operators are
\begin{equation}
X(g)=\proj{ +}+\proj{ -} e^{i\alpha_g},
\end{equation}
while the $Z$ operator reads 
\begin{equation}
Z=\ket{+}\bra{-}.
\end{equation} 
This specific model can be mapped (on bipartite lattices) by a unitary transformation to the one studied in  \cite{orland_exact_1992}.

\begin{figure}[!hbt]
\includegraphics[height=5cm]{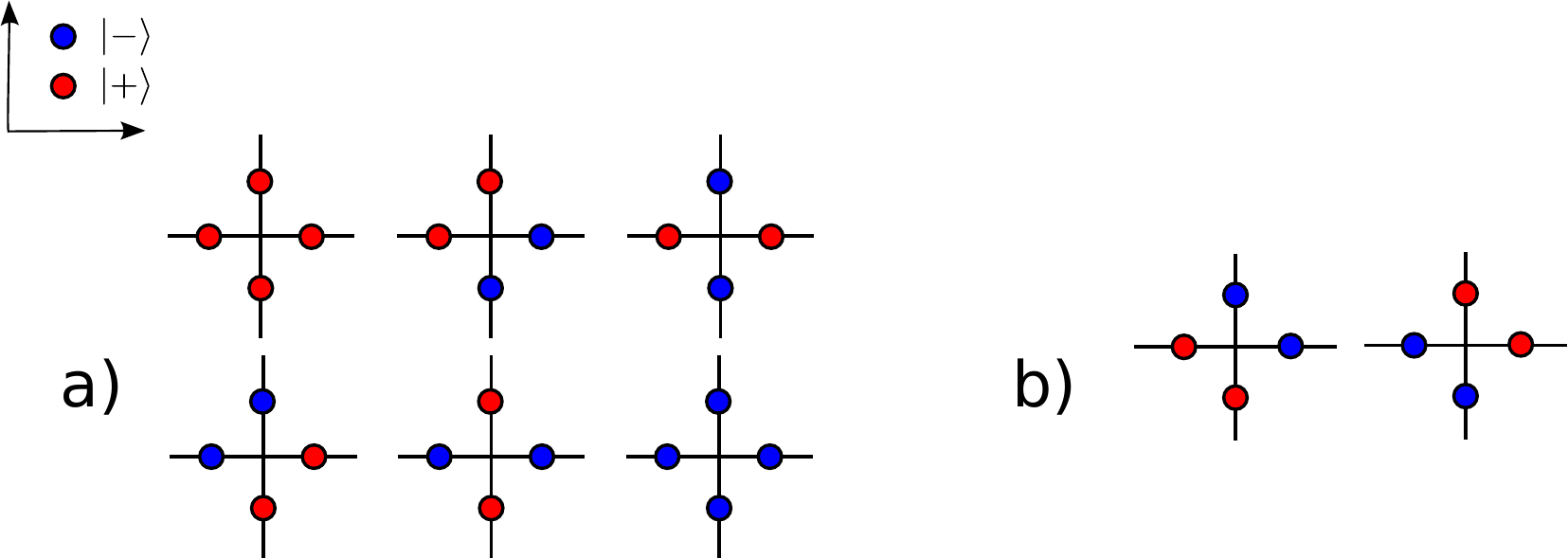}
\caption{{\bf a)} The requirement of gauge invariance of the states using   (\ref{eq:gauge_cond})  selects only 6 out of the total 16 states of the four qubits entering a given site. In the standard 2D orientation, one can visualize the 6 allowed states by coloring in red states $\ket{+}$ and blue $\ket{-}$. \label{fig:allowed_configurations} {\bf b)} If we consider the standard  $\mathbb{Z}_2$ LGT  the gauge symmetry condition of   (\ref{eq:gauge_cond}) selects $8$ out of the $16$ states of the four qubits attached to the links entering a given site. They include the same $6$ states than those of the U$(1)$ gauge magnet of panel a). The  two extra states are represented here, and are those responsible of the existence short closed loops.} 
\end{figure}

 In order to describe the physics of the model, it is important to notice that   we can  characterize ${\cal H}_G$ graphically. In the $\ket +$ and $\ket -$ basis and with  the standard 2D orientation
of Fig. \ref{fig:def}, the gauge condition (\ref{eq:gauge_cond}) on a given site $s$ selects among the $2^{4}$
states of the four links entering $s$ only six possible configurations
\begin{equation}
\ket{+,+,+,+},\ket{-,+,-,+},\ket{+,-,+,-},\ket{-,-,-,-},\ket{-,+,+,-},\ket{+,-,-,+}.\label{u1tginvstate}
\end{equation}
These are shown in Fig. \ref{fig:allowed_configurations}. Similarly to standard LGT, we can interpret the U$(1)$-invariant Hilbert space as the space of closed strings formed by  $\ket{-}$ states, onto the vacuum of $\ket{+}$ states. Indeed, as any of the six allowed states contains an even number of $\ket{-}$, at each site of the lattice, for any ingoing $\ket{-}$ there is also the corresponding outgoing one. Hence, the string cannot end on a (bulk) site, i.e. only closed string (or string touching the boundary) are compatible with gauge invariance. 
In addition,  it is worth to notice that the six allowed configurations do not allow to construct a closed string of finite size (without reaching the boundary). This means that, for lattices with periodic boundary conditions,  
strings can only close by wrapping around the whole lattice. This implies that their length is at least equal to the lattice size. This is an important difference with respect to standard LGT.
 In standard LGT, indeed, closed strings can be of arbitrary length, the shortest being the  strings around a single plaquette.

 In order to clarify the origin of this discrepancy, one can consider the Kogut-Susskind $\mathbb{Z}_2$ LGT introduced in the previous section. There, the group algebra is two-dimensional, that is, the local Hilbert space is still made of qubits.
 As we have already shown, the gauge condition selects $8$ out of the $16$ states  of the four links entering a given site. It turns out that $6$ of them coincide with the one of the $U(1)$ gauge magnet  of Fig. \ref{fig:allowed_configurations} a), but there are  two extra states 
\begin{equation}
\ket{-,-,+,+},\ket{+,+,-,-},
\end{equation}
as shown in Fig. \ref{fig:allowed_configurations} b).

\begin{figure}[!hbt]
 \includegraphics[height= 5cm]{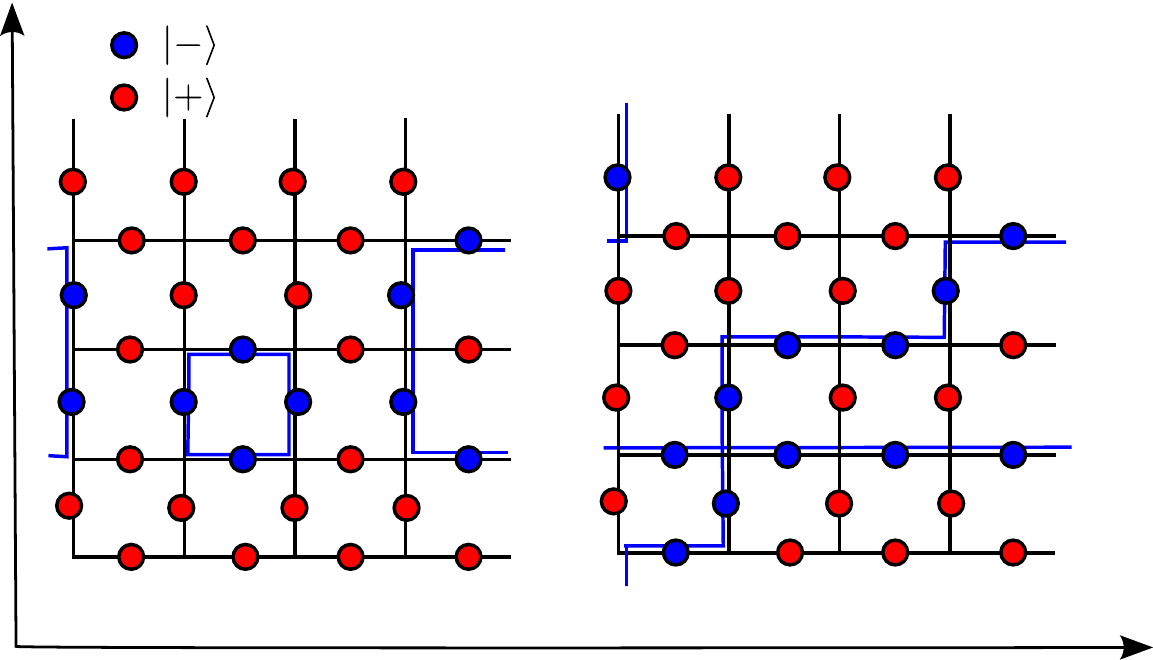}
\caption{ Here, we depict a 4x4 lattice with periodic boundary conditions in both directions. Red qubits are in the state $\ket{+}$, while blue qubits are in the state $\ket{-}$. The blue qubits form closed string pattern as required by gauge invariance. On the left, we show a possible gauge invariant state of the  standard $\mathbb{Z}_2$ LGT. On the right, the same plot is presented for the U$(1)$ gauge magnet. In both cases, the states can be mapped to closed string configurations. There is an important difference between the two cases, however. In $\mathbb{Z}_2$ LGT,  closed strings can form arbitrary small closed loops. In the U$(1)$ gauge magnet, due to the absence of the states in Fig. \ref{fig:allowed_configurations} b), strings are forced to close  by wrapping around the whole  lattice. This means that their typical length exceeds that of the linear lattice size. Blue lines are drawn as a guide to the eye to recognize the closed string wrapping around the periodic boundaries of the lattice. \label{fig:strings}}
\end{figure}

These states are, indeed, the ones needed to close a small loop of $\ket{-}$ (for example a single plaquette). In fact, this small difference produces completely different strings patterns, as we can appreciate in Fig. \ref{fig:strings}.

The different string pattern have deep consequences on the physics of the models.
Standard LGT, indeed, have at least two phases, a confined phase and a deconfined phase. In the confined phase, short closed string abound, while long closed string are very rare and vice-versa in the deconfined phase  \cite{levin_string-net_2005}.
In the $U(1)$ gauge magnet, by just noticing the absence of short closed strings, we already have a strong indication that the phase diagram of the model is very  different  from the one of the standard $U(1)$ LGT.

\begin{figure}[!hbt]
 \includegraphics[height=5cm]{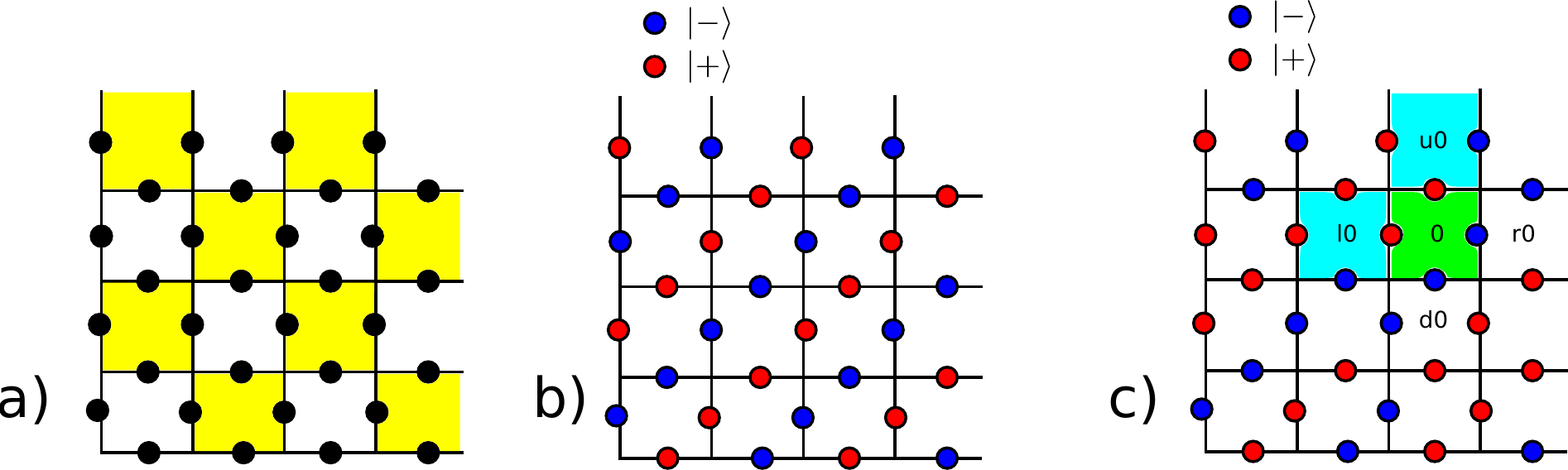}
\caption{{\bf a)} Lattices with an even number of sites and  periodic boundary conditions are bipartite. This allow to consider the model with  half of the  plaquettes (filled in yellow in the figure) turned on. In this way, the  different yellow plaquettes do not  share any link and the  U$(1)$ gauge magnet Hamiltonian is built up of mutually commuting terms. This implies that  the ground-state is simultaneous eigenvector of all the terms appearing in the Hamiltonian. {\bf b)} Reference state used to construct the ground-state of the Hamiltonian (\ref{h0}) through the action of projectors (\ref{eq:gs_projector}). {\bf c)} An example of reference state used to obtain an excitation with gap $\Delta=2$. \label{fig:bipart}}
\end{figure}

Let us analyze it in details.
First, let us consider the dynamics induced by the Hamiltonian (\ref{eq:ham_ks}) when $\theta =0$.
Contrary to what happens in standard Abelian LGT, for the gauge magnets two plaquettes operator sharing one link are not commuting, i.e., generally  $[B_p,B_p']\neq 0$, . This means that the ground-state is not simultaneous eigenvector of all the $B_p$'s. However, for the sake of the present discussion we can, as suggested in  \cite{orland_exact_1992}, circumvent this problem on bipartite lattices by considering, first, only half of Hamiltonian, and, afterwards, the effects of the other terms in (\ref{eq:ham_ks}).  
We start with
\begin{equation}
 H_0 = -\sum_{p_y} (B_{p_y}+B_{p_y}^\dagger),\label{h0}
 \end{equation}
where with $p_y$ we label half of the plaquettes, the ones drawn in  yellow in Fig. \ref{fig:bipart}. In this way,  $H_0$ only contains operators that do not share any link, and, thus, all commuting.  

The model (\ref{h0}) is exactly solvable, and we can write its ground-state as the action of a series of projectors onto a given reference state. It is important to notice that the reference state should be i)  gauge invariant, and ii) not belonging to the kernel of the various $B_p$ considered. Indeed, in contrast to what happens for standard LGT, the $B_p$ here  has only two eigenvalues different from zero, equal to $\pm 1$. It is easy to check that a possible reference state fulfilling these requirements is the one depicted in the central panel of  Fig. \ref{fig:bipart},  consisting of two strings each of length $L^2/2$ wrapping around the  lattice of size $L \times L$ with periodic boundary conditions. We call this reference state  $\ket{\psi_0}$. We can define the three different projectors on to the subspace of different eigenvectors of $(B_p+B_p^\dagger)$ as
\begin{equation}
 P^{1}_p= \frac 12(B_p+B_p^{\dagger})\left( B_p+B_p^{\dagger} +  \mathbb{1}\right) \label{eq:gs_projector},
\end{equation}
\begin{equation}
 P^{-1}_p= \frac 12(B_p+B_p^{\dagger})\left( B_p+B_p^{\dagger} -  \mathbb{1}\right) \label{eq:m1_projector},
\end{equation}
\begin{equation}
 P^{0}_p= - \left(B_p+B_p^{\dagger} + \mathbb{1}\right) \left(B_p+B_p^{\dagger} - \mathbb{1}\right)\label{eq:0_projector},
\end{equation}
so that the ground-state $\ket{\Omega(0)}$ is proportional to
\begin{equation}
 \ket{\Omega(0)} \propto \bigotimes_{p_y} P^{1}_{p_y} \ket{\psi_0}.\label{eq:gs}
\end{equation}
By construction the above state is not null and minimizes the energy as 
\begin{equation}
  \ket{\Omega(0)} = \bigotimes_{p_y} \ket{1}_{p_y}, \quad \left(B_{p_y}+B_{p_y}^{\dagger} \right) \ket{1}_{p_y}= \ket{1}_{p_y}, \forall p_y\in \text{\{yellow plaquettes\}.} \label{sp1}
  \end{equation}

As, in principle, all the eigenvectors of $H_0$ can be constructed by applying the projectors (\ref{eq:gs_projector}-\ref{eq:0_projector}) on an appropriate gauge invariant state, it follows that there is a gap to the first excited state. Such gap is at least equal to one, corresponding to a state
\begin{equation}
\ket{\Phi(\{0,p'_y\})} = \bigotimes_{p_y\neq p'_y}\ket{1}_{p_y}\otimes\ket{0}_{p'_y}, \quad \left(B_{p'_y}+B_{p'_y}^{\dagger} \right) \ket{0}_{p'_y}= 0.\label{sp0}
\end{equation}
 However, it is easy to argue that such a state cannot satisfy the gauge invariance requirement on each of the sites at the vertices of the plaquette $p'_y$. Hence, $\ket{\Omega(0)}$ describes  a gapped phase with gap $2$. The simplest excited state with such energy is 
 \begin{equation}
 \ket{\Phi(\{-1,p'_y\})} = \bigotimes_{p_y\neq p'_y}\ket{1}_{p_y}\otimes\ket{-1}_{p'_y}, \quad \left(B_{p'_y}+B_{p'_y}^{\dagger} \right) \ket{-1}_{p'_y}= -\ket{-1}_{p'_y},\label{sp-1}
\end{equation}
as $ \ket{\Phi(\{-1,p'_y\})} \propto \bigotimes_{p_y\neq p'_y} P^1_{p_y}\otimes P^{-1}_{p'_y} \ket{\psi_0}$. 
It is worth to notice that first excited states of the form $\ket{\Phi(\{0,p'_y\},\{0,p^{\prime\prime}_y\})} = \bigotimes_{p_y\neq p'_y,p^{\prime\prime}_y}\ket{1}_{p_y}\otimes\ket{0}_{p'_y}\otimes\ket{0}_{p^{\prime\prime}_y}$ also exist and are gauge invariant, but cannot be obtained just by applying projectors on the reference state $\ket{\psi_0}$ as $P^0_{p_y} \ket{\psi_0}= 0$, $\forall p_y$ in the yellow plaquettes' sublattice. In this case, the procedure is more complex and it involves a $B$ operator acting on a plaquette of the other sublattice, as illustrated in the right pannel of Fig. \ref{fig:bipart}. It is found that
\begin{equation}
\ket{\Phi(\{0,l_0\},\{0,u_0\})} \propto P^0_{d_0} P^0_{r_0} B_0\ket{\Omega(0)},
\end{equation}
where $d_0,r_0,u_0,l_0$ are the plaquettes in the yellow sublattice that surrounds the $0$ plaquette of the complementary sublattice.
 
In this simple model, we can compute exactly how the presence of static external charges modifies the ground-state.
A static   charge  $\pm 1$ at site $s$ modifies the gauge condition (\ref{eq:gauge_cond}) to 
\begin{equation}
 A_s(g)\ket{\psi}=\exp(\pm i\alpha_g) \ket{\psi}. \label{eq:charges}
\end{equation}
That is to say, the  presence of static charges  modifies the allowed string configurations. Open strings are, indeed, allowed if  they start and end on the charges. This, clearly, affects the properties of the ground-state of the system.
With two charges $\pm 1$, the ground-state is orthogonal to the one without static charges.
This implies that we have to change both the reference state   and the set of projectors, in order to construct the new ground-state
\begin{equation}
 \ket{\Omega(0)_{\pm}} \propto \bigotimes_{p_y} \tilde{P}_{p_y} \ket{\psi_{\pm1}}.\label{eq:gs_charge}
\end{equation}

\begin{figure}
 \includegraphics[width=10cm]{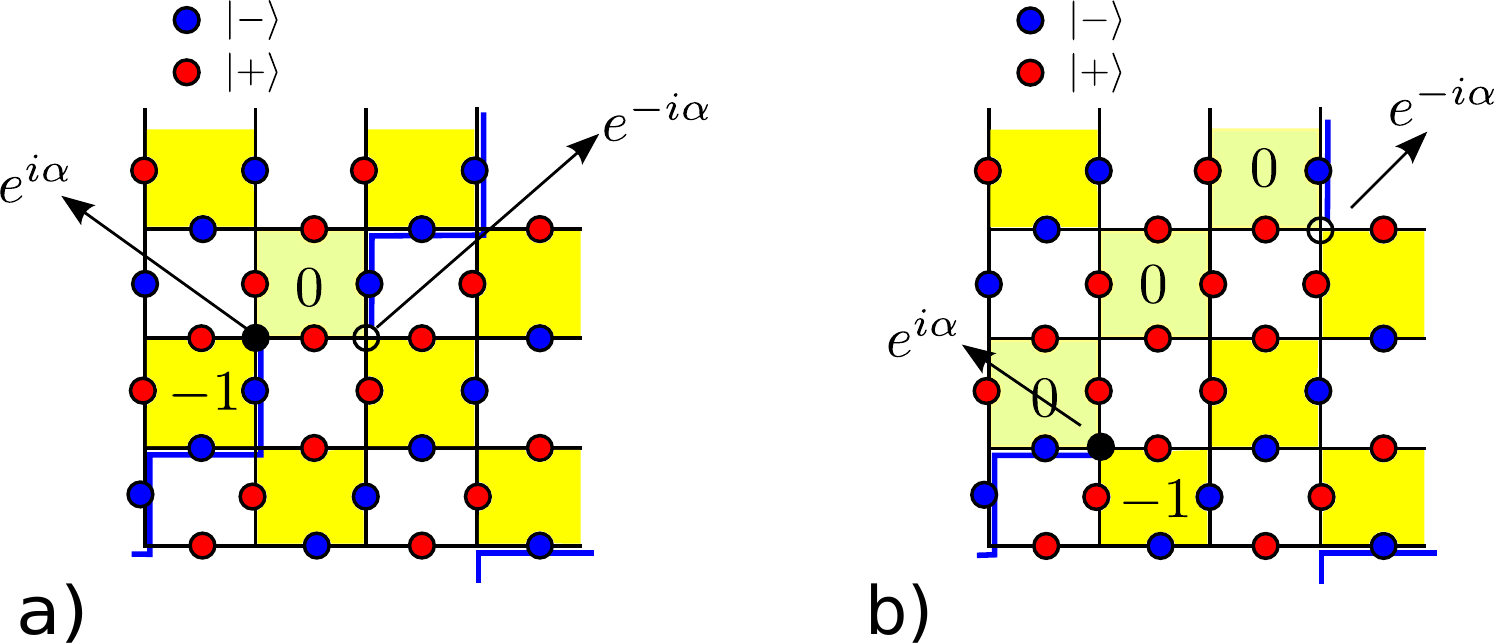}
\caption{The presence of static external charges modifies the ground-state. {\bf a)} Sketch of the reference configuration used to obtain the ground-state with two static first neighbors  charges (presented as filled black dots for charge $+1$ and empty dots for charge $-1$). The blue string os states $\ket{-}$ is broken in between them. The plaquettes states along the broken part of the string are orthogonal to the  eigenstates of $B_p$ with eigenvalue $1$. These plaquette are filled in with a pale shade of yellow  in the drawing  while regular plaquettes that contribute $-1$ to the ground-state energy filled in solid yellow.  This implies that the ground-state energy in the presence of static charges increases by an amount proportional to the inter-charge separation, manifestation  of charge confinement. {\bf b)} Sketch of the reference state for charges  at  distance $r=2\sqrt{2}$. The reference state induces a  further pale yellow plaquette with respect to the reference state  on the left and has thus higher energy. 
 \label{fig:stat}}
 \end{figure}

A simple candidate  for $\ket{\psi_{\pm1}}$ can be obtained by transforming one of the two closed strings of $\ket{-}$ contained in $\ket{\psi_0}$ into  an open one. As illustrated in in Fig. \ref{fig:stat} for a $4 \times 4$ lattice with periodic boundary condition, this can be done by flipping one or more consecutive $\ket{-}$ (blue) links to (red) $\ket{+}$ links.  This creates  two static charges of opposite charge at the two ends of the blue string, where modified gauge condition (\ref{eq:charges}) holds. Two possible choices of the  reference state $\psi_{\pm1}$, which differ on the  position of  the static charges, are sketched  in Fig. \ref{fig:stat}. The $+1$ static charges are denoted by filled black dots, while  $-1$ ones  by empty dots.  

Now, the projectors $\tilde{P}_{p_y}$ are determined by the requirement of minimizing the energy while respecting the new gauge condition, which depends only on the position of the $\pm 1$ charges. Since the above defined $\ket{\psi_{\pm1}}$ differ from $\ket{\psi_0}$ only by few flipped links, located between the two ends of the open string, the projectors $\tilde{P}_{p_y}$ may be distinct from the $P^1_{p_y}$ only for the plaquettes $p_y$ (of the yellow lattice) interested by such flips, $p_y\in\{$flipped region$\}$ (pale yellow shaded plaquettes of Fig. \ref{fig:stat}).  In fact, as manifestation of orthogonality of the  ground-state with back-ground charges  to the ground-state without charges, $\ket{\psi_{\pm1}}$ is annihilated by the projectors (\ref{eq:gs_projector}) acting on the  plaquettes between the two ends of the open string. 
Hence, the proper choice of projectors in such a region, which minimizes the energy and does not annihilate the state, is given by the projector of (\ref{eq:0_projector}), $\tilde{P}_{p_y}=P^0_{p_y}$, $\forall p_y\in\{$flipped region$\}$. 

As a consequence, the ground-state energy of the system with two static charges is higher than the one  without the two charges. The energy gap $\Delta_q$ is equal to the number of $P^0$ projectors, i.e., the number of yellow plaquettes that contain flipped links. Up to artifacts of the discretization, such number is proportional to the number of flipped links itself, $n_F$, i.e., is proportional to the distance $r$, in lattice spacing unit, between the  charges, $\Delta_q(r)\propto r$. As it can be easily deduced from Fig. \ref{fig:stat}, the exact relation between the gap and the number of flipped links is 
$$\Delta_q = n_F - \text{Int}[\frac{n_F}2],$$
while the distance between the charge depends on $n_F$ as
$$r=\left|\left(n_F - \text{Int}[\frac{n_F}2] \right) (1,0) + \text{Int}[\frac {n_F}2 ] (0,1)\right|= \sqrt{n_F^2 + 2\; {\text{Int}[\frac {n_F}2 ]}^2 - 2\; n_F \,\text{Int}[\frac {n_F}2 ]}  ,$$
where Int$[x]$ is the integer part of $x$.  
The above formula is obtained by taking in account the ``zig-zag'' behavior the displacement of the charges shows when one additional link is flipped.


These two relations together are the footprint of charge confinement, a  phenomenon that the present model shares with many other gauge theories, among them, QCD.
This means that the gap scales as 
\begin{equation}
 \Delta_q(r) = \sigma r,\label{eq:string_tens}
\end{equation}
where $\sigma$ is called string tension.
In the long distance regime, i.e., $n_F\gg1$, it follows that $\Delta_q\sim \frac{n_F}2$ and $r\sim \frac { n_F}{\sqrt 2}$, hence, $\Delta_q\sim \frac r{\sqrt 2} $, 
and the string tension for this model is $\sigma\sim \frac 1{\sqrt{2}}$.  

At this point, we consider the effect of the link term for $\theta\neq 0$. By the Hamiltonian  $\cos \theta\, H_0 + \sin \theta\, \sum_l(X_l(g)+ X^\dagger_l(g))$. The  last term of the Hamiltonian, $\cos \theta\, H_0 + \sin \theta\, \sum_l(X_l(g)+ X^\dagger_l(g))$ is certainly dominant close to $\theta=\frac{\pi}{2}$. In this phase, the ground-state is the product state
 \begin{equation}
  \ket{\Omega\left(\frac{\pi}{2}\right)} = \bigotimes_l \ket{-}_l,
 \end{equation}
regardless of the $\alpha_g$ we have chosen to represent  a generic  $U(1)$ element $g$. Hence, one expects to encounter  a phase transition when $\theta$ grows from zero to $\frac{\pi}{2}$. We observe that i) 
\begin{equation}
H_0\ket{\Omega\left(\frac{\pi}{2}\right)} = 0, \label{cond1}
\end{equation}
i.e., the expectation value of the plaquette part of the Hamiltonian vanishes in the ground-state at $\theta=\frac{\pi}{2}$, and, as well,
ii)
\begin{equation}
\sum_{l\in p}(X_l+X_l^\dagger) \ket{1}_p = 4(1+\cos \alpha_g)\ket{1}_p, \forall p,\label{cond2}
\end{equation}
which implies $\sum_l (X_l+X_l^\dagger)\ket{\Omega(0)} = (1+\cos \alpha_g) L^2 \ket{\Omega(0)}$. From i) and ii), it immediately follows that the states $\ket{\Omega(0)}$  and   $\ket{\Omega\left(\frac{\pi}{2}\right)}$ are not deformed, but simply shifted in energy, by the change of $\theta$. Hence, the transition between them is a first order  phase transition (level crossing) where the expectation value of any (yellow) plaquette operator can be taken as an order parameter which jumps abruptly from $-1$  to $0$.

Furthermore, the relations (\ref{cond1},\ref{cond2}) allow to compute the critical value of the coupling $\theta_c$ simply by equating the energy per plaquette of $\ket{\Omega(0)}$  and   $\ket{\Omega\left(\frac{\pi}{2}\right)}$ 
\begin{equation} 
 \theta_c : -\cos \theta_c + 4 \sin \theta_c(1 + \cos \alpha_g)  =  8  \sin(\theta_c) \cos(\alpha_g),\, \to \theta_c= \arctan\left(\frac{1-\cos(\alpha_g)}4\right).
\end{equation}

Also in the product state phase, static charges are confined. The minimal configuration containing to two charge excitations of opposite sign, $\pm 1$, is  given by the shortest open string  of $\ket{+}$-links connecting the charges.
These configurations have an energy cost per link that is constant and equal to $2-2\cos(\alpha_g)$. Again, the linear behavior of energy gap with the charge distance is a manifestation of the confinement of charges.

Finally, we are ready to study the whole Hamiltonian (\ref{eq:ham_ks}). We notice that although the actual shape of ground-state in the plaquette dominated phase $\ket{\Omega(0)}$ is modified, the conditions (\ref{cond1}-\ref{cond2}) are not, as the plaquette ground-state has always an equal number of plus and minus links. This implies just a change in the actual value of $\theta_c$, but not in the nature of the phase transition that remains a first order level-crossing phase transition.

Hence, all the properties of the plaquette dominated phase can be studied at $\theta=0$, as a change of $\theta$ only induces   a shift in the energy of such state and of the states associated to plaquette excitations. 

\begin{figure}
\includegraphics[width=8cm]{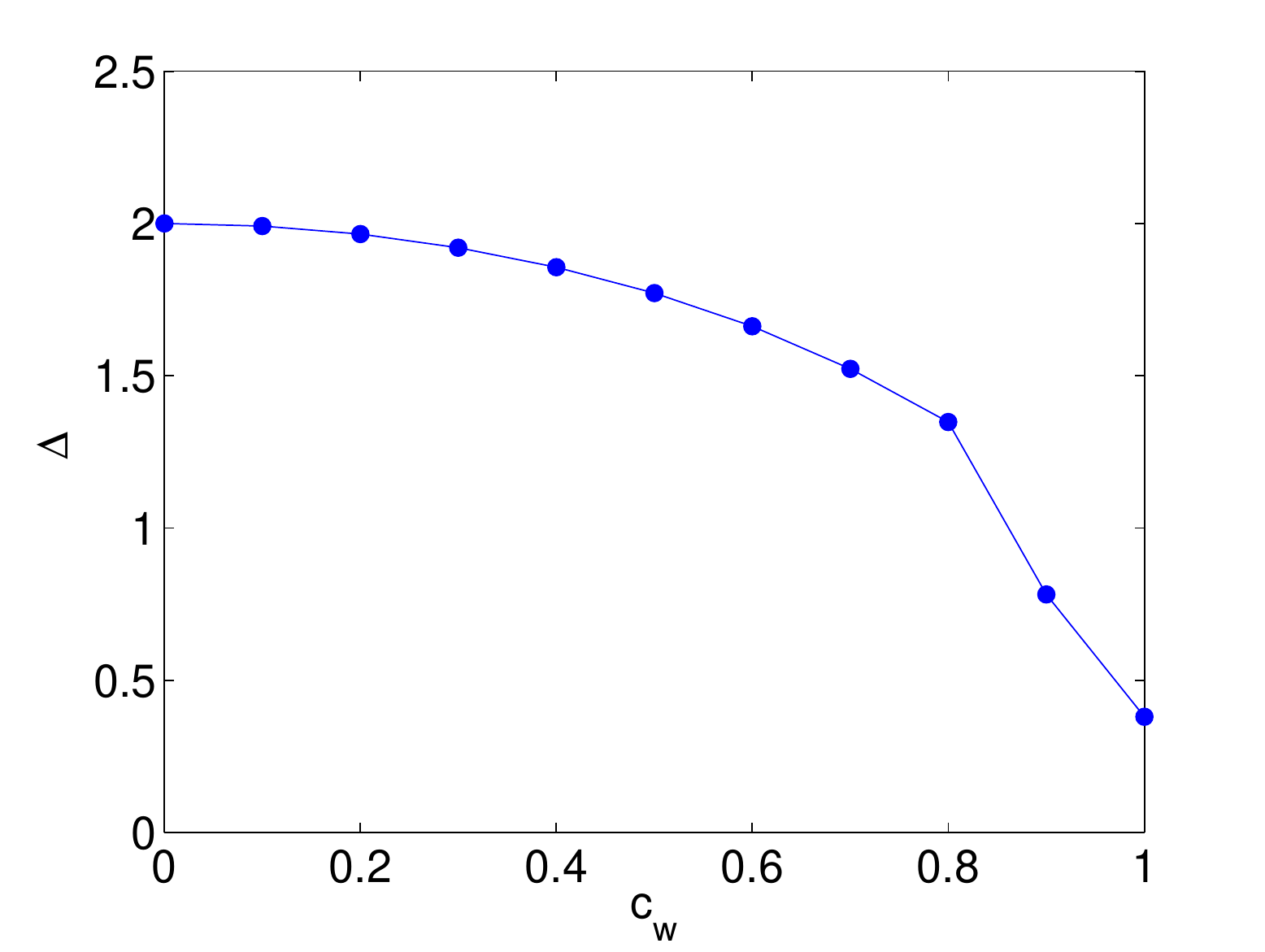}
  \caption{Scaling of the gap $\Delta$ between the ground-state and the first excited state of the Hamiltonian of  (\ref{eq:ham_w}) as a function of  the coupling $c_w$ for a $4 \times 4$ lattice with periodic boundary conditions in all directions. The gap decreases and eventually saturates to its finite size value. On a infinite lattice the results of  \cite{orland_exact_1992}\label{fig:gap} predict that it would vanishes at the point at $c_w=1$.}
 \end{figure}

In order to obtain the full gauge magnet Hamiltonian, we can add the other half of the plaquettes to the Hamiltonian (\ref{h0}) adiabatically
\begin{equation}
 H_{c_w} = -\sum_{p_y} (B_{p_y}+B_{p_y}^\dagger) -c_w \sum_{p_w} (B_{p_w}+B_{p_w}^\dagger) \label{eq:ham_w}
\end{equation}
where $p_w$ are the white plaquettes in Fig. \ref{fig:bipart}. By varying $c_w$ from $0$ to $1$ the system  is driven to a gap-less phase, \cite{orland_exact_1992}. The dependence of the gap on $c_w$ for the $4\times 4$ lattice is shown in Fig. \ref{fig:gap}. There we appreciate that the gap systematically decreases and eventually saturates to its finite size value.

We conclude the section by making a comparison between the regular and the two-dimensional representation of U$(1)$ LGT. The truncated formulation has the clear advantage of being easier to handle numerically, and to simulate in optical lattice.  
The relevant question is whether the truncated and the non-truncated theories share the same quantum properties. It is worth to notice that the gapless phase encountered here has no analogue in the standard compact U$(1)$ LGT,  where the only appearing phase is  the gapped confining phase  \cite{polyakov_compact_1975}. 

There are several interesting physical phenomena that are however common to both models. Here, we just list two of them, which we propose in the following section as possible test to check whether the desired physics is correctly  realized in the quantum simulator.
The first  one is related to an important theorem that states that gauge symmetries can not be broken spontaneously \cite{kogut_introduction_1979}. The physical consequence of this fact is that only  
gauge invariant operators can acquire a non-vanishing expectation value. This implies that the expectation value of, i.e., a single $Z$ operator on a link has to be zero, property that can be easily checked with current experimental resources.
The second one is related to charge confinement.  In the $U(1)$ link model, we have seen that both phases exhibit charge confinement. This is similar to the charge confinement encountered in the Kogut-Susskind formulation of compact $U(1)$ LGT. As discussed in section \ref{veri}, the presence of charge confinement  can be verified experimentally.

Definitely, in order to establish the correct relation, beyond analogies,  among the standard compact QED and the present link model -- including a full  understanding of  the  renormalization group  flow of the U$(1)$ gauge magnet--  further studies are needed.

\subsection{Exact diagonalization and Tree Tensor Network of the U$(1)$ gauge model on  small lattices}\label{4b4}
\begin{figure}
\includegraphics[width=13cm]{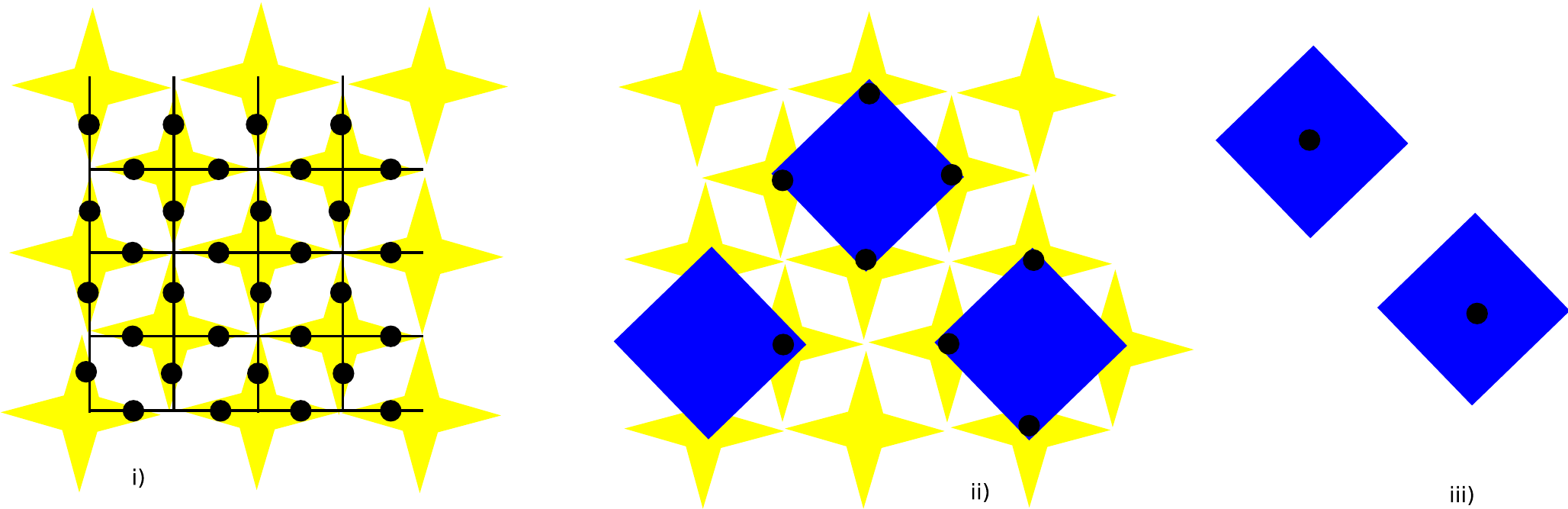}
 \caption{In order to perform exact diagonalization on the $4 \times 4$ lattice, we need to appropriately reduce the number of degrees of freedom by using the constraints induced by gauge invariance. There are several way to perform the reduction. Here, we introduce an RG inspired procedure that can be generalized to larger lattices and would immediately provide a tree tensor network ansatz for symmetric states (see also \cite{tagliacozzo_entanglement_2011}). The procedure consists in  identifying  blocks of four spins and project only on those configurations selected by the gauge conditions (those  sketched in Fig. \ref{fig:allowed_configurations}).  This produces a new lattice of effective sites with dimension $6$. We can iterate the procedure and obtain a final layer where only few spins are left. At this point, the effective Hamiltonian  can be diagonalized directly. If one needs to consider larger lattices, the projection onto gauge invariant states has to be followed by a projection onto low-energy 
states, 
 such that one can iterate a larger number of time the RG steps. See, e.g., \cite{tagliacozzo_entanglement_2011,tagliacozzo_simulation_2009}.\label{fig:rg_gauge} }
\end{figure}

With a specific choice of  open boundary conditions the model is exactly solvable. Here, we performed an exact diagonalization of a system with periodic boundary conditions in both directions to see which of the features of the exactly solvable model  survive the presence of periodic boundary conditions.
Clearly, with this approach we are limited to study small systems, up to $4 \times 4$. For this reason, we also explain how one can go beyond that, by using the idea presented in  \cite{tagliacozzo_entanglement_2011}. The gauge conditions are, indeed, fixed in a RG inspired way, so to reduce the number of degrees of freedom. We can use a $4 \to 1$ coarse graining scheme as follow. We cover the lattice with independent $A_s$ by chosing one of each two crosses and perform a RG step by keeping only the corresponding $6$ eigenvectors with eigenvalue 1 of the corresponding $A_s$, from the available $16$ states. This reduces the $2^32$ degree of freedom of the $4\times4$ lattice to only $8^6$. The $8$ remaining crosses still act on four coarsegrained sites. Now, we can repeat the procedure by fixing two further crosses and select among the $6^4$ states the corresponding $6 \times 3^4$.
In this way, we can perform exact diagonalization on the two effective spins. The unused gauge constraints are projected at this level. This procedure is sketched in Fig. \ref{fig:rg_gauge}.
If we want to deal with systems larger than $4 \times 4$ we cannot  use the above construction. However, the introduction of an addtional RG step, which projects onto the low-energy subspace of the full Hamiltoniana, allows to study  systems up to $10 \times 10$ (see, e.g., \cite{tagliacozzo_simulation_2009}). This can be done numerically as described in detail in  \cite{tagliacozzo_entanglement_2011}.

\section{Quantum simulation of the U$(1)$ gauge magnet}\label{sect:ryd}

In this section, we discuss the quantum simulation of Abelian, in particular U$(1)$, gauge magnets in optical lattices. As the gauge magnets may be regarded as  a  prototype of lattice gauge theories, many of the major issues of the implementation are generic to any gauge theory.

\subsection{Many-body interaction through the Rydberg gate}

\begin{figure}
 \includegraphics[height=5cm]{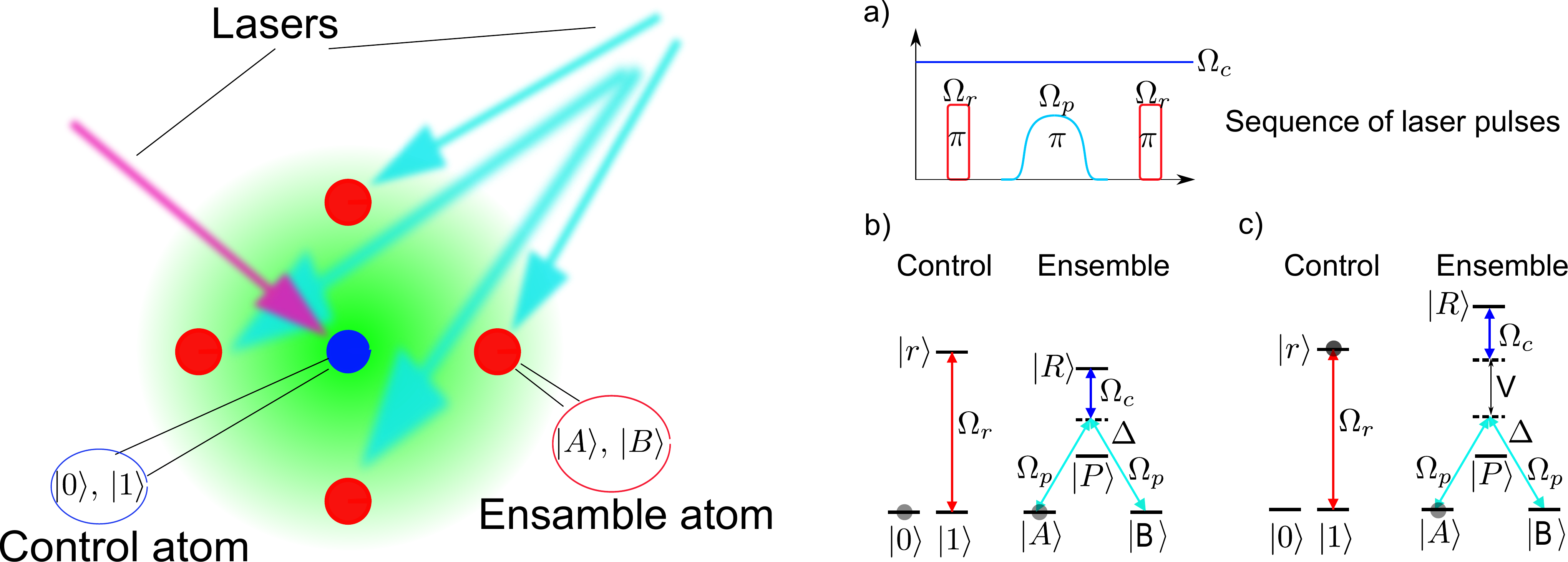}
\caption{Schematic of the Rydberg gate. In the block of the left, the geometric disposition of the ensemble and control atoms is shown. The ensemble atoms, which admit as logical states $\ket A$ and $\ket B$, are placed sufficiently close to the control atom, whom logic states are $\ket 0$ and $\ket 1$, so that the Rydberg interaction, i.e., the dipole interaction in the Rydberg state of the control, is strong  (shadow green volume). At the same time, the atoms are sufficiently separated to allow individual single-site addressing by laser pulses. In the block of the right, the pulses, needed to the conditional transfer from the logical state $\ket A$ to the logical state $\ket B$ of the ensemble atoms, and the level scheme are displayed. a) Sequence of the pulses (not in scale). The $\pi$-pulse  of Rabi frequency $\Omega_r$ (in red) is applied on the control atom {\it only} and induces resonant transition between level $\ket 1$ and the Rydberg excited state $\ket r$. Such pulse is short and applied {\it twice}, at the beginning and at the end of conditional transfer. The pulse of Rabi frequency $\Omega_c$ (in blue) and the $\pi$-pulse of Rabi frequency $\Omega_p$ (in cyan) act both on all the ensemble atoms simultaneously. The former is shining continuously and much more intense than the latter, which is smoothly ramped up and down. b) Level scheme for the control initialized in $\ket 0$. In this case, the pulse $\Omega_r$ does not play any role. The $\pi$-pulse $\Omega_p$ connects with an (off-resonant) two-photon transition between the states $\ket A$ and $\ket B$ of the ensemble atoms. The frequency of the pulse $\Omega_c$ is such that $\Omega_c+\Omega_p$ connect the logic states $\ket A$ and $\ket B$ to the Rydberg excited state $\ket R$ with a resonant two phon transition. As $\Omega_c\gg\Omega_p$, the Electromagnetic induced transparency scheme (EIT) is realized and the transfer between $\ket A$ and $\ket B$ inhibited. c) Level scheme for the control initialized in $\ket 1$. In this case, the pulse $\Omega_r$ excites the control to the Rydberg state $\ket r$. Due to dipole interaction the Rydberg state of the ensemble $\ket R$ gets shifted in energy by $V$ and it is no longer in a two-photon resonant transition with the logical state $\ket A$ and $\ket B$. Hence, the transfer between $\ket A$ and $\ket B$ occurs.\label{fig:ryd} }
\end{figure}

An important feature common to gauge theories on a generic lattice is the occurrence of interactions between several links together, involving at least as many links as the lattice coordination number. As such links appear in the dual lattice as sites, it implies the existence of a long range interaction between several sites at once. The aforementioned feature is very hard to be simulated. For instance, it is  very hard to engineer in a controlled way such an interaction between cold atoms loaded in optical lattices. Such problem disfavors the analogic simulation of lattice gauge theory, i.e., the implementation of the gauge theory Hamiltonian as the effective Hamiltonian of the controlled system. 

The  alternative is to consider a digital quantum simulation, i.e., the simulation of time-evolution of gauge theories via quantum gates. Its realization requires the capability of assigning a certain phase to a set of links,phases determined by the expectation value of each multilink operator entering the  Hamiltonian of interest. The gates that implement these operations are generalizations of C-NOT gates, and the key role of control over links at distance is played by Rydberg atoms \cite{rydberg}.

Rydberg atoms are highly excited states that are long lived and have a huge dipolar moment \cite{blockade1,blockade2}. The latter gives raise to the phenomenon known as blockade: once a Rydberg atom has been laser induced, a volume of several $\mu$m is affected by its dipole field causing level shifts in the surrounding atoms, and, hence, forbidding the creation of  another Rydberg excitation in such volume.  Several experiments have demonstrated that it is  possible to create entanglement over an extended region in a (laser) controlled way by exploiting such strong dipolar interaction \cite{rydberg_experiments}.

Indeed, due to the blockade, the presence/absence of a Rydberg excitation plays the role of an ideal control qubit, as it may trigger or forbid laser stimulated transitions in all the atoms in the blockade volume, for instance all the atoms trapped in a unit cell of a lattice. The explicit realization of this idea has been pushed forward in  \cite{Muller09}, where the authors propose the mesoscopic Rydberg
 gate scheme, which is sketched in Fig. \ref{fig:ryd}. 
 Let us review briefly the main feature of this gate: for more details and for the analysis of the fidelity we refer the reader to the original reference.  
 Let us consider a set of $N+1$ qubits, one of which playing the role of the control. The logic states of the control $\{ \ket 0 ,\ket 1\}$, and of the remaining ensemble qubits, $\{ \ket{A}_k ,\ket{B}_k \}$, $k=1,\dots,N$, are hyperfine states of atoms. The goal is to achieve a protocol such that the ensemble states are untouched when the control is in $\ket 0$ while induces a flip $\ket{A}_k\leftrightarrow\ket{B}_k$ when the control is in $\ket 1$. In formula, this reads
 \begin{equation}
 G=\ket 0 \bra 0 \otimes \mathbb{1}_N +  \ket 1 \bra 1 \otimes \left(\bigotimes_{k=1}^N \sigma^x_k\right),\label{g}
 \end{equation}
 where $\mathbb{1}_N$ is the identity in the space of the ensemble qubits and $\sigma^x$ is the Pauli matrix acting as $\sigma^x_k \ket{A}_k = \ket{B}_k$, $\sigma^x_k \ket{B}_k = \ket{A}_k$, $\forall k=1, \dots, N$.
  We may suppose that both the control and the ensemble atoms are trapped in different sites of an optical lattice and that we have single site addressing. All the atoms can be excited by laser pulses to Rydberg states, which we indicate as $\ket r$ and ${\ket R}_k$ for the control and the ensemble atoms, respectively. $\ket r$ and ${\ket R}_k$ can be in principle different.  The key idea is to use a  Electromagnetic Induced Transparency (EIT) \cite{eit} in a modified Lambda scheme.  Three different $\pi$-pulses are employed with Rabi frequency  $\Omega_r$, $\Omega_c$ and $\Omega_p$, respectively. The first one is applied to the control at the beginning and  at the end of the process and it is resonant between $\ket 1$ and $\ket r$. This implies that the control is excited to the Rydberg state only when it is in the logic state $\ket 1$.  The other two lasers are such that the sum of their frequencies is exactly resonant with the transition ${\ket A}_k,{\ket B}_k\to {\ket R}_k$ when the control atom is not 
in the Rydberg state $\ket r$, while it is off resonant when the strong dipole-dipole interaction is present (a more accurate discussion should take into account also the effect of simultaneous Rydberg excitations in the ensemble atom as done originally in  \cite{Muller09}; such extra-dipole interactions can be kept under control and  the functionality of the gate is not spoiled). Together with the condition $\Delta\gg\Omega_c\gg\Omega_p$, where $\Delta$ is the smallest detuning of the $p$-laser from an intermediate atomic level $\ket P$, the above construction ensures that no transfer between ${\ket A}_k$ and ${\ket B}_k$ occurs when they are in two-photons resonance with ${\ket R}_k$, i.e., the control is in $\ket 0$; on the contrary  the transfer occurs with efficiency one in the absence of the two-photons resonance i.e., the control is in $\ket 1$. This ideal result is not spoiled once the losses due to spontaneous decay of the long-live atomic states are taken into account.

  Before moving to the applications of such gate to the simulation of the  gauge magnets, two comments are in  order.
First, the mesoscopic Rydberg gate can be modified to get any $G'$ of the form
\begin{equation}
G'=\ket 0 \bra 0 \otimes \mathbb{1}_N +  \ket 1 \bra 1 \otimes \left(\bigotimes_{k=1}^N e^{i\frac{\alpha_g}2 \hat n_k \cdot \boldsymbol{\sigma}_k}\right)= \bigotimes_{k=1}^N U_k(\hat n_k)\left(\ket 0 \bra 0 \otimes \mathbb{1}_N +  \ket 1 \bra 1 \otimes \left(\bigotimes_{k=1}^N e^{i\frac{\alpha_g}2 \sigma^x_k}\right)\right)\bigotimes_{k=1}^N U_k^\dagger(\hat n_k) ,\label{ggen}
 \end{equation}
where $\alpha_g$ is any real number between 0 and $2 \pi$, $\hat n_k$ a generic 3-vector of modulus 1, $\boldsymbol{\sigma}\equiv (\sigma_x,\sigma_y,\sigma_z)$, and $U_k(\hat n_k)$ the unitary transformations in each single qubit space. Indeed, the $\alpha_g$ dependence is simple obtained by considering an $\alpha_g$-pulse for the lasers $c$ and $p$ instead of the $\pi$-pulse considered originally in the mesoscopic Rydberg gate. On the other hand, in order to get the $\sigma$ combination $\hat n_k \cdot \boldsymbol{\sigma}$, the single site addressability is exploited to perform the appropriate rotation of each qubit basis. This operation obviously amounts to apply  single qubit gates on the ensemble atoms before and after the $G$ gate.

Second, as originally shown in  \cite{Weimer10}, the mesoscopic Rydberg gate 
can be used to read out the expectation of a (two-valued) operator on the ensemble atoms and transfer the information to the control qubit, once combined with rotations on the control itself. Indeed, let us consider the effect of the unitary transformation
\begin{equation}
\tilde G = \exp[i\frac{\pi}4\sigma^y_c]\, G\, \exp[-i\frac{\pi}4\sigma^y_c],\label{readergate}
\end{equation}
once the control qubit is initialized in $\ket 0$. For a generic state of the ensemble atoms $\ket \lambda = \alpha_g \ket{\lambda_+} + \sqrt{1-\alpha_g^2} \ket{\lambda_-}$, where with $\ket{\lambda_\pm}$ are eigenvectors of $O_e\equiv \bigotimes_k \sigma^x_k$, $O_e \ket{\lambda_\pm}=\pm \ket{\lambda_\pm}$, and  $|\alpha_g|<1$, the final state of the control is computed to be $\ket c =  \alpha_g \ket{0} + \sqrt{1-\alpha_g^2} \ket{1}$. Furthermore, by inverting $\tilde G$, any single qubit operation applied to the control can be mapped to the ensemble state, i.e., on the logical qubit composed by $\ket{\lambda_+}$ and  $\ket{\lambda_-}$. In case we are interested to read a more complicated operator, as in (\ref{ggen}), appropriate rotations of the ensemble qubits are required as well, $\tilde G'=  \exp[i\frac{\pi}4\sigma^y_c]\, G'\, \exp[-i\frac{\pi}4\sigma^y_c]$.

\subsection{Dissipative preparation ${\cal H}_G$ as ground-state preparation of frustration-free Hamiltonian }\label{diss}

\begin{figure}
\includegraphics[width=9cm]{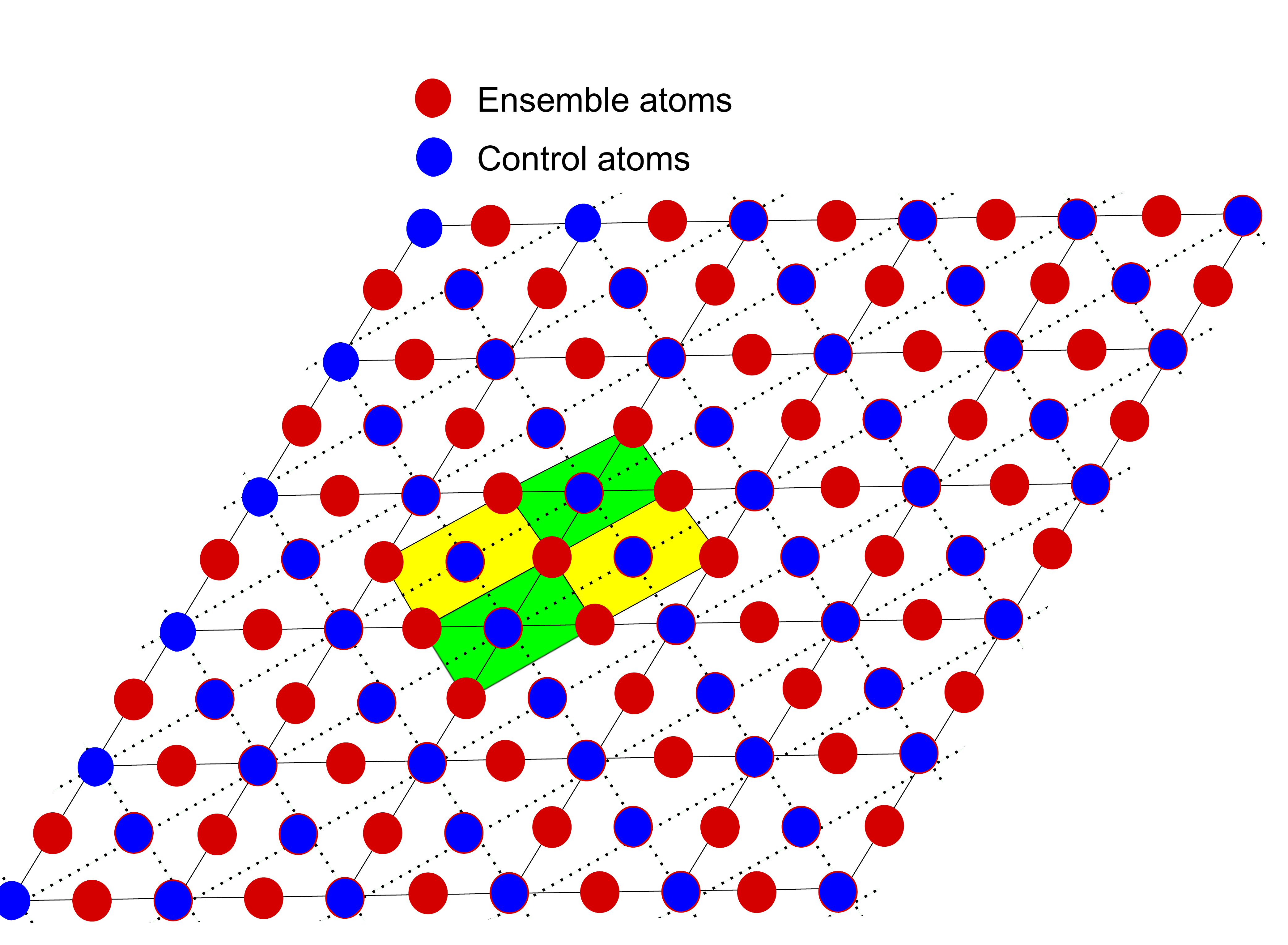}
  \caption{Schematic of the simulator. As indicated in the legend, the red dots represent the ensemble atoms, while the blue dots represent the control ones. Between the control atoms, we distinguish the ones at the lattice sites (intersections of continuous lines) and the ones at the center of the plaquettes. The former are needed to impose the gauge condition and interact with the ensemble atoms at the corners of the green squares. The latter control the dynamics of the plaquettes, which involves the ensemble atoms at the corner of the yellows squares.  
   \label{fig:simulator}}
 \end{figure}

We are now ready to discuss in full detail the protocol for the simulation of the gauge magnets. The link qubits are identified with the logical states of the ensemble atoms (see Fig. \ref{fig:simulator}). For convenience, we choose the following relation between the $\{\ket{+},\ket{-}\}$ link basis of section \ref{truu1} and the logical (atomic level) basis $\{\ket{A},\ket{B}\}$
\begin{equation}
\ket{\pm}=\frac{\ket{A}\pm\ket{B}}{\sqrt 2}.
\end{equation} 
The first conceptual and practical step is to impose the gauge condition (\ref{eq:gauge_cond}), i.e., to reduce the possible states of the system to the gauge invariant ones. It is worth to notice that this is equivalent to find the (highly degenerate) ground-state of the Hamiltonian $H_{cond}=-\frac 12\sum_s (A_s (g)+A^\dagger_s (g))$, with $g$ a generic element of $\mathbb{Z}_N$ or of U$(1)$ such that $g^n\ne e$ for $n<N$, or for any $n$, respectively. In the U$(1)$ case, we may choose $H_{cond}$ to be
\begin{equation}
H_{cond} \equiv \sum_s h^s_{cond}= -\frac 12\sum_s e^{i\frac{\alpha_g}2 \sigma^x_{s_1}} \otimes e^{i\frac{\alpha_g}2 \sigma^x_{s_2}} \otimes e^{-i\frac{\alpha_g}2 \sigma^x_{s_3}}\otimes e^{-i\frac{\alpha_g}2 \sigma^x_{s_4}} +H.c. , \label{hcond}
\end{equation}
where $\alpha_g$ is any integer, for instance 1. 

Independently of the specific election for $H_{cond}$, it is, evidently, a frustration-free Hamiltonian: each local term commutes with the others, and the ground-state can be obtained by minimizing each local term separately.  
In this situation, a dissipative preparation of the ground-state, as originally proposed in  \cite{kraus_preparation_2008,Verstraete09}, and applied to quantum simulation in  \cite{Weimer10} (for a very recent review see \cite{Muller12}), is appropriate. The idea of dissipative preparation is very simple and corresponds to a classical cooling. That is to say that each time the local state has no minimal energy is transformed to a minimal energy eigenstate by a local operation. 
Let us detail the procedure in the case of (\ref{hcond}). Our aim is to read out the expectation of $A_s$ ($h^s_{cond}$) on the local state and to change it, for instance by flipping an ensemble qubit, if is not satisfying the gauge condition (or equivalently $\langle h^s_{cond}\rangle =-1$). As our local state is given by the four links attached to a site and $A_s$ is a unitary operation that can be conditionally implemented (see discussion above) by the mesoscopic Rydberg gate, the required operation is obtained with the gate sequence
\begin{equation}
U_D = \tilde G(A_s)^{\dagger} U_s^{z,\varphi} \tilde G(A_s),\label{UD}
\end{equation}
where   $\tilde G(A_s)= e^{i\frac {\pi}4 \sigma_c^y}(\proj{0} \otimes \mathbb{1}_s + \proj{1}\otimes A_s)e^{-i \frac{\pi}4 \sigma_c^y}= \frac 12 \mathbb{1}_c\otimes(\mathbb{1}_s+ A_s) + \frac 12 \sigma_c^x\otimes (A_s -\mathbb{1}_s)$, and 
$$U_s^{z,\varphi}= \proj{0}\otimes \mathbb{1}_s + \proj{1} \otimes \Sigma(\varphi),$$
is the conditional flipping of one of the links entering $s$ with probability $\varphi$, let say $s_1$, $\Sigma(\varphi)=e^{i\varphi \sigma_{s_1}^z}$. 

Two comments are in order. First, we note that the spin flip $\Sigma(\varphi)$ is not necessary projecting the local state $\ket{\psi}_s$ to the subspace $\langle A_s\rangle=1 $ in one step, even for $\varphi=\pi/2$. Furthermore, if $h^s_{cond} \ket{\psi}_s = -\cos (2\alpha_g) \ket{\psi}_s$, it follows that $h^s_{cond} \Sigma(\frac {\pi}2)\ket{\psi}_s= -\cos (\alpha_g) \Sigma(\frac {\pi}2)\ket{\psi}_s$ but if $h^s_{cond} \ket{\psi}_s = -\cos (\alpha_g) \ket{\psi}_s$, it is found $\Sigma(\frac {\pi}2)\ket{\psi}_s$ has component on the $-1$ and $-\cos (2\alpha_g)$ subspace with amplitudes $\sqrt{\frac 34}$ and $\frac 12$ 
\cite{Note2}. That is to say, the ``energy'' is not monotonically decreasing at 
each step.  However, the relation above is telling us that, by iterating the process, the probability of the state to be excited is exponentially suppressed, as the probability of being reheated scales like $(\frac 14)^{n/2}$, $n\equiv$ \# of steps.

Second, it is worth to notice that the operation $U_D$ is unitary, but the dissipation enters in the game as, in order to iterate $U_D$, we need to reset the control qubit to ${\ket 0}_c$. As explained in  \cite{Weimer10}, such decoupling of the control dynamics can be obtained by inducing a radiative decay of ${\ket 1}_c$. This implies that the final result of one dissipative iteration is a mixed state. Indeed, by naming $\ket{\psi}_s$ any initial pure local state and assuming that the control qubit is initialized in $\ket 0$, the action of $U_D$ is
\begin{equation}
\ket 0 \otimes \ket{\psi}_s \xrightarrow{U_D} \ket 0 \otimes \ket{\phi_0}_s + \ket 1 \otimes \ket{\phi_1}_s,
\end{equation}
where 
\begin{equation}
\ket{\phi_0}_s= \frac 12\left(1 + \frac{A_s+A_s^\dagger}2 + \frac{(A^\dagger_s - \mathbb{1}_s)\Sigma(\varphi)(A_s - \mathbb{1}_s)}2\right)\ket{\psi}_s,\label{phi0}
\end{equation}
\begin{equation}
\ket{\phi_1}_s= \frac 12\left(\frac{A_s^\dagger - A_s}2 + \frac{(A^\dagger_s + \mathbb{1}_s)\Sigma(\varphi)(A_s - \mathbb{1}_s)}2\right)\ket{\psi}_s.\label{phi1}
\end{equation}
 This means that, in consequence of the decay of $\ket{1}$ to $\ket{0}$, the density matrix after a dissipation step evolves as
 \begin{equation}
\ket 0 \otimes \proj{\psi}_s \xrightarrow{\text{Dissipation}} \proj{\phi_0}_s + \proj{\phi_1}_s.
\end{equation}
 The most efficient cooling procedure is for $\varphi=\frac{\pi}2$. However, as discussed in  \cite{Weimer10}, it is possible to show that, in the limit of small $\varphi$, the above cooling procedure is governed by a master equation of the Lindblad form. By defining the jump operators 
 \begin{equation}
 c_0\equiv -\frac{(\mathbb{1}- A_s^\dagger)\sigma_1^z (1-A_s)}4=c_0^\dagger,\ \ \ \ \ \ \ c_1\equiv \frac{(\mathbb{1}+ A_s^\dagger)\sigma_1^z (1-A_s)}4,
 \end{equation}
 the expansion of (\ref{phi0}) and (\ref{phi1}) in $\varphi$ around $\varphi=0$ takes the form
  \begin{equation}
 \ket{\phi_0}_s= \left(\mathbb{1} -i \varphi c_0 - \frac{\varphi^2}2 (c_0^2+c_1^\dagger c_1) + O(\varphi^3)\right)\ket{\psi}_s,\ \ \ \ \ \ \  \ket{\phi_1}_s= \left(-i \varphi c_1  + O(\varphi^2)\right)\ket{\psi}_s,
 \end{equation}
 which implies that
the variation of $\rho$ after one cooling step can be written up to second order in $\varphi$ as 
\begin{equation}
\Delta \rho= \Delta t \, \dot \rho = -i\varphi [c_0,\rho] + \varphi^2 \left( c_0 \rho c_0 + c_1 \rho c_1^\dagger - \frac 12 \{(c_0^2+c_1^\dagger c_1),\rho\} \right) + O(\varphi^3),
\end{equation}
 where 	$\Delta t$ is the time needed for a the single time step. It is immediate to see that the above equation has no other fixed points than a density matrix built up of pure states,  $\rho=\sum_j \proj{\psi_j}$, satisfying $A_s\ket{\psi_j}=\ket{\psi_j},\forall j$.  In comparison to \cite{Weimer10}, we find that two jump operators are associated to the cooling procedure and the  presence of a coherent term linear in $\varphi$, which is not affecting the convergence as it is zero on the gauge invariant subspace.

\subsection{Unitary evolution of the U$(1)$ gauge magnet}\label{unievo}

Once the gauge condition has been successfully implemented, we can focus on the dynamics of U$(1)$ gauge magnet, governed by the Hamiltonian (\ref{eq:ham_ks}). As explained above, our goal is to implement the unitary evolution $U_t=e^{-iH t}$ in terms of local unitary gates \cite{Weimer10}. The simplest way to do it is to exploit the Trotter expansion to write
\begin{equation}
U_t\sim \prod_p \left(e^{-i H_p \Delta t} e^{- i( \sum_l H_l) \Delta t}\right)^{\frac t{\Delta t}},
\end{equation}
where $H_p$ and $H_l$ are the plaquette, $H_p= -\cos\theta (B_p+B_p^\dagger)$, and the link $H_l= \sin\theta (X_l+X_l^\dagger)$ terms, respectively. As the Rydberg gate are sufficiently fast ($\Delta t\simeq \mu s$) compared with temperatures achieved for the atoms in the lattice, the above approximation is consistent and we can treat each term as commuting in the time-evolution. 

Let us focus first in the implementation of the link part. As the $X_l$ commute between them, such term is just a product of the single qubit rotation $U_l= e^{-i (1+\cos \alpha_g)t}e^{-i(1-\cos\alpha_g) t \sigma_l^x}$ applied to each link. Thus, its implementation is straightforward.

On the other hand, the plaquette term is slightly more involved. The easiest way to treat this term is to expand explicitly the exponential into a sum of  tensor product of $\sigma$. In order to reduce the number of terms to be simulated, we may use the Trotter expansion further. As $Z$ and $Z^\dagger$ can be written as $Z=\frac{\sigma^z-i\sigma^y}2$ and $Z^\dagger=\frac{\sigma^z + i\sigma^y}2$, respectively, we have
\begin{eqnarray}
H_p &\equiv&-\frac{\cos\theta}8\sum_{j=1}^8 Q_p^{(j)}\cr
    &   =& -\frac{\cos\theta}8 \left((\sigma^z)^{\otimes 4} + (\sigma^y)^{\otimes 4} + (\sigma^z\otimes\sigma^y)^{\otimes 2} + (\sigma^y\otimes\sigma^z)^{\otimes 2} + \sigma^z\otimes\sigma^y\otimes\sigma^y\otimes\sigma^z + \sigma^y\otimes\sigma^z\otimes\sigma^z\otimes\sigma^y \right.\cr
    &  &\hspace{3cm}\left. -(\sigma^z)^{\otimes 2}\otimes(\sigma^y)^{\otimes 2} -(\sigma^y)^{\otimes 2}\otimes(\sigma^z)^{\otimes 2}\right),\label{pladeco}
\end{eqnarray}    
which gives
\begin{equation}
e^{-i H_p \Delta t} \sim \prod_{j=1}^8 e^{i \frac{\cos\theta}8 Q_p^{(j)}\Delta t}.
\end{equation}    
Now, each of the above unitaries can be implemented using Rydberg gates as explained in  \cite{Weimer10}, by applying the sequence
\begin{equation}
e^{i \frac{\cos\theta}8 Q_p^{(j)}\Delta t}=\tilde G(Q_p^{(j)})^\dagger e^{i \frac{\cos\theta}8 \Delta t \sigma_c^z}\tilde G(Q_p^{(j)}),
\end{equation} 
for $\tilde G(Q_p^{(j)})= \exp[i\frac{\pi}4\sigma^y_c] ( \proj{0}\otimes\mathbb{1}_p + \proj{1}\otimes  Q_p^{(j)}
) \exp[-i\frac{\pi}4\sigma^y_c]$, which is of the form of (\ref{ggen}). 

It is worth to conclude with a final remark on the preservation of gauge invariance during the above process.  In principle, as the Hamiltonian commutes with the gauge condition (\ref{eq:gauge_cond}), the time-evolution conserves the gauge condition, i.e., it is not driving the system out of the gauge invariance subspace. In fact, in order to minimizing possible violations of the gauge condition due to dephasing it is worth to keep applying the dissipative process  while other unitary operations are performed. By doing so, the unitary evolution is not altered.

\subsection{Proposal for the preparation of the U$(1)$ gauge magnet ground-state}\label{sec:statepreparation}

We are finally ready to discuss the preparation of the U$(1)$ gauge magnet ground-state. As in the link dominated phase the ground-state is a fully polarized product state that can be efficiently prepared using single site addressing, we focus on the plaquette dominated phase. Our plan is to realize in practice the procedure explained in section  \ref{truu1}. Hence, the idea is to construct the ground-state in two steps.  
First, 
we apply the dissipative procedure to the frustration-free Hamiltonian obtained considering half of the plaquette (for instance the yellow ones in Fig. \ref{fig:bipart}) in order to find the corresponding ground-state $\ket{\Omega(0)}$ defined by (\ref{eq:gs}) and (\ref{sp1}), and, then, 
we drive such state,  adiabatically or using the CRAB procedure \cite{doria_optimal_2011}, to the ground-state of the full plaquette Hamiltonian. It is worth to notice that prior to the ground-state preparation it is not necessary to impose the gauge condition as $\ket{\Omega(0)}$ is gauge invariant by construction. However, for the driving, the final remark
 of the previous section applies. 
 
 The procedure explained above works for any chosen background of static charged matter, once the proper initial state is prepared and the right gauge condition is applied at site. As in section \ref{truu1}, we focus on 1) the preparation of the ground-state in the absence of charges and on 2) the preparation of the ground-state in presence of  two isolated + and - charges.
 
 \subsubsection{Preparation of the ground-state in the absence of charges}   

The preparation of the initial state to be driven can be obtained  by applying the discussion of section \ref{diss} to the local Hamiltonian $H_{p}$, $p\in$\{yellow plaquettes\}. We note that the unitary operator  $U_{H_p}\equiv \exp[i\frac {\pi}2 (B_p + B_p^\dagger)]$ acts on the eigenvector of $H_p$ (see definitions (\ref{sp1}-\ref{sp-1})) as
\begin{equation}
   U_{H_p} \ket{1}_p=\ket{1}_p,\ \ \ \ \ \ \  U_{H_p} \ket{-1}_p=- \ket{-1}_p,\ \ \ \ \ \ \  U_{H_p} \ket{0}_p=-i \ket{0}_p,
\end{equation}
i.e., in a very similar way to the star operator $A_s$. That is to say, we can get the right dissipative process just by substituting $A_s$  with $U_{H_p}$ in the gate derived in section \ref{diss}. However, the C-NOT gate $\proj{0}\otimes \mathbb{1}_p + \proj{1}\otimes U_{H_p}$ is hard to be implemented exactly (i.e., without applying Trotter approximation) using Rydberg gates. 

We can bypass easily such difficulty by exploiting the exact knowledge of $\ket{\Omega(0)}=\bigotimes_{p\in\{\text{Yell.pl.}\}} \ket{1}_{p}$, where the local state $\ket{1}_{p}$ can be explicitly written by ordering the links of plaquette anticlockwise starting from the bottom as 
\begin{equation}
\ket{1}_{p}=\frac 1{\sqrt 2} \left(\ket{+,+,-,-}_p + \ket{-,-,+,+}_p\right).
\end{equation}
 We observe that such a state is completely determined by: i) being a combination of states with an equal number of $+$ and $-$, ii) being the stabilizer state of both $O^x\equiv\sigma^x\otimes\sigma^x\otimes\mathbb{1}\otimes\mathbb{1}$ and $O^z\equiv\sigma^z\otimes\sigma^z\otimes\sigma^z\otimes\sigma^z$ i.e., satisfying  $O^\mu \ket{1}_{p}=\ket{1}_{p}$, $\mu=x,z$. 
 
 The condition i) can be realized in different ways. As suggested in section  \ref{truu1}, the most direct one is to prepare the product state $\ket{\psi_0}=\bigotimes_{p\in\{\text{Yell.pl.}\}} \ket{+,+,-,-}_{p}$ by exploiting single addressing.
As an alternative,  we could first impose the gauge condition and, then, repeat the dissipative procedure of section \ref{diss} by choosing as local state the (yellow) plaquettes, as unitary operator $A_p={\exp[i\frac{\pi}4 \sigma^x]}^{\otimes 4}$, and as flip operator the simultaneous rotation of two plaquette's qubit by $\sigma^z$, $\Sigma=e^{i\varphi\sigma_1^z}\otimes e^{i\varphi\sigma_2^z}$. Indeed, the gauge condition ensures that the initial plaquette state has an even number of $+$ and the double flip converts states with zero o four $+$ in a desired state.

The condition ii), once i) is realized, can be achieved by applying the Kraus maps for Bell state pumping engineered in  \cite{Barreiro11}. Indeed, let us suppose that we have prepared the state $\ket{\psi_0}$. Hence,  we have to  impose by dissipation only $\langle O^z \rangle=1$ on the plaquette state as $O^x\ket{\psi_0}=1$ by construction. The dissipative gate
$$ \tilde G(O^z)^\dagger U^{z,\varphi=\frac {\pi}2}\tilde G(O^z) + \text{radiative decay} \ket{1}\to\ket{0}$$
does the job in one step. 
 
Now, we are ready to turn on the remaining plaquette interactions. This can be done by increasing the coupling constant $c_w$ in (\ref{eq:ham_w}) {\it adiabatically} from 0 to 1. From the digital simulation point of view, this means that unitary time-evolution procedure of section \ref{unievo} has applied to the plaquette Hamiltonian (\ref{eq:ham_w}),
where $c_w$ is changed smoothly, for instance linearly, from an iteration to an other. 
However, as first shown in  \cite{orland_exact_1992} and reviewed in section \ref{truu1}, in the infinite dimensional system, the plaquette dominated phase becomes gapless as $c_w$ approaches 1, with the gap that scales as $\frac 1{c_w^2}$. This means that the adiabatic driving for a finite system is still possible but it becomes harder and harder as the dimension of the lattice increases. This implies that the ramping time has to be taken bigger and bigger. In order to reduce such time, optimal control techniques can be used. Between them, the so call CRAB method \cite{caneva_chopped_2011}, which has been successfully employed, for instance, to dramatically reduce the time needed to bring a superfluid gas into a Mott insulator state \cite{doria_optimal_2011}, seems very promising as it allows for closed-loop optimization experiments. The idea is to find the optimal function $c_w(t)$, satisfying $c_w(0)=0$, such that after a time $T$ minimizes a cost function ${\cal F}$, which in our case is a linear 
combination of the energy of the driven state computed with the final Hamiltonian $c_w=1$ and the {\it fluences}, a measure of the oscillation of $c_w(t)$. The novel feature of the CRAB method is that the optimal driving at the step $I$, $c^{(I)}_w(t)$, is obtained by minimizing ${\cal F}$ with respect to few free parameters $p_k$ in the driving probes $c^{(I-1)}_w(t) \sum p_k g_k(t)$. The $g_k(t)$ are element of a favorite  orthonormal basis, for instance the Fourier basis $g_k(t)=\cos(k\frac{\pi}T t)$. In order to further improve the convergence, the frequencies $w_k=k\frac{\pi}T$ may be randomly perturbed $w_k\to w_k + r_k$. One then 
feeds the result of the optimization back to the algorithm so that the final energy is improved further up to convergence.

\begin{figure}
\includegraphics[width=8cm]{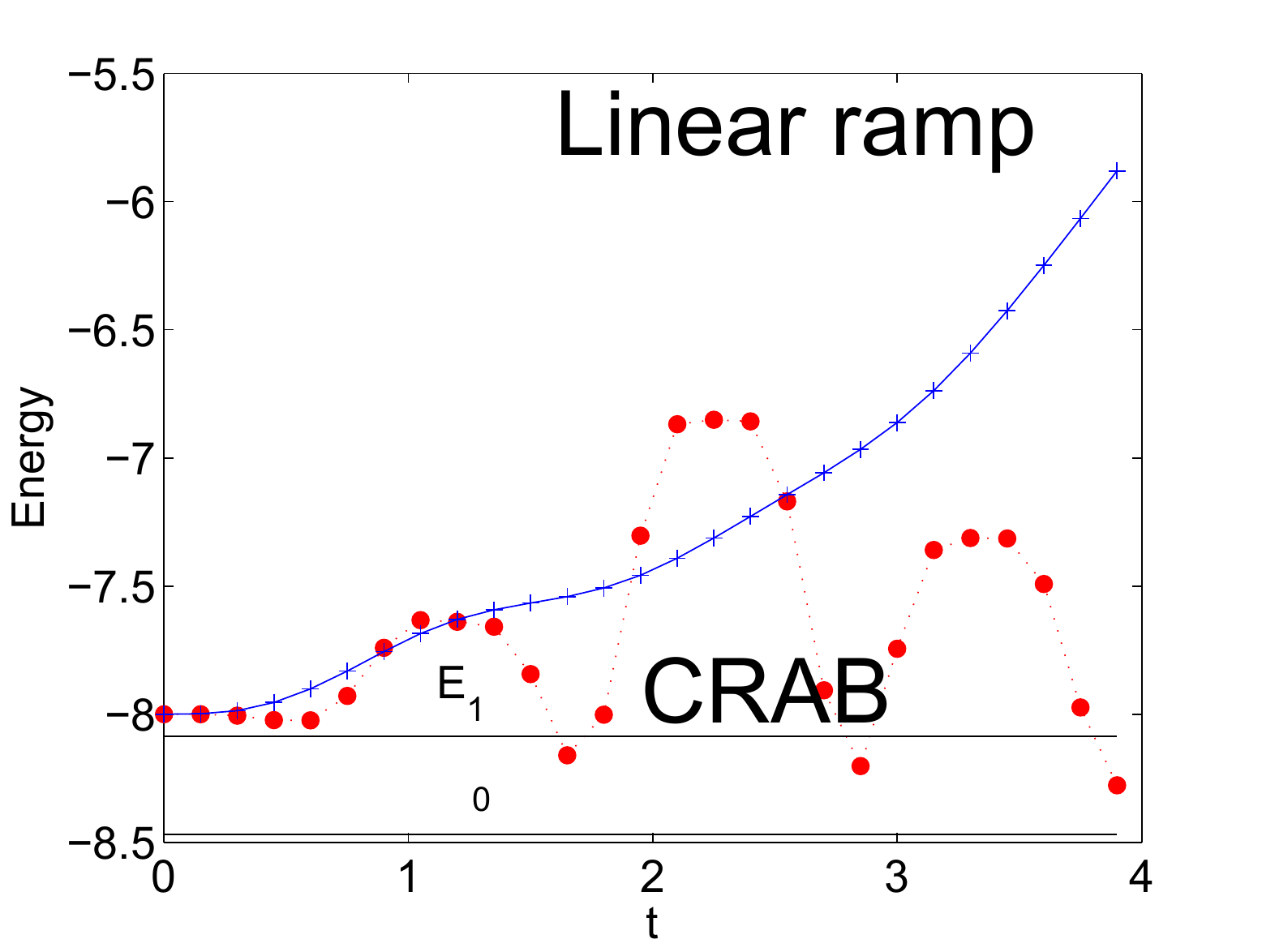}
 \caption{Comparison between linear and  CRAB protocol drivings.  The preparation of the ground-state of the gauge magnet Hamiltonian (\ref{eq:ham_w}) is performed in a given time, shorter than the adiabatic one. The blue line represents a linear ramp from $c_w=0$ to $c_w=1$. The red curve represents the result of one optimization cycle of a CRAB preparation with seven random frequencies as free parameters \cite{caneva_chopped_2011}. The horizontal lines represent the energies of the ground-state and the first excited state. It can be appreciated how the linear  map fails completely to deliver the desired ground-state while the CRAB map gets very closed to it.\label{fig:crab}}
\end{figure}

We have numerically tested the CRAB method up to a $4\times 4$ lattice, where we are still able to compute the exact ground-state for $c_w=1$ using exact diagonalization. One has to keep in mind that  the energy of the final state of the time-evolution  is  monotonically decreasing with the number of iterations. Since each iteration takes very long (typically around a week on a standard PC) we have decided to stop our optimization algorithm as soon as the final energy got well below the energy of the first excited state. This typically happens after only one optimization cycle. For the sake of comparison we report in Fig. \ref{fig:crab} the energy of the state along its time-evolution and compare it with respect to the energy of the state obtained using  a linear ramp that goes from $c_w=0$ to $c_w=1$, in the same amount of time. 
It can be clearly seen how out of the adiabatic regime the linear ramp fails completely to
prepare the ground-state, as the evolution is by far too fast, while the CRAB optimization with only one iteration gets very closed to it.

\subsubsection{Preparation of the ground-state in presence of  two isolated + and - charges}

In presence of two isolated + and - charges, two main difference in the preparation occur. First, the initial state to be driven, i.e. the ground-state of the system with only half of the plaquettes turned on (the yellow sublattice of Fig. \ref{fig:bipart}),  is different. Indeed, as explained in section \ref{truu1}, it is again a product state of eigenvectors of the plaquettes turned on, but it differs from the ground-state in the absence of charges on few plaquettes, which we called  {\it flipped}. On the flipped plaquettes, gauge invariance forces to take an eigenvector of eigenvalue 0. Furthermore, the flipped plaquettes can only occupy a continuous segment of a southeast-northwest diagonal of the sublattice, see Fig. \ref{fig:stat}. Accordingly, the charges can be placed consistently with gauge invariance only on the sites of the first and last plaquette of the segment.
 To be precise, a charge can occupy all the sites except the one shared with the contiguous flipped plaquette  of the interior. Hence, by definition the flipped plaquettes of the interior contains no charges. The position of the charges fully determines which are states on the flipped plaquettes.
 Let assume for convenience that the positive charge is on the first flipped plaquette at the bottom-left and the negative one is on the last at the top-right. It follows that  the proper state for the plaquettes in the interior of the segment is 
 $\ket{++++}$. For the plaquettes on the frontier we have two possibility: the charge occupies the site, $d$, on the diagonal of the segment (not shared with other flipped plaquettes), or it occupies one of the two off diagonal. In the former case the state is again  $\ket{++++}$. In the latter, the link connecting $d$ with the site of the charge is in $\ket -$ with all the other links are in $\ket +$. Hence, it follows that the plaquette at the bottom-left is in the state $\ket {-+++}$ if the charge is below the diagonal or in $\ket{+++-}$ if the charge is above. Similarly,  the plaquette at the top-right is in the state $\ket {+-++}$ if the charge is below the diagonal or in $\ket{++-+}$ if the charge is above.  Such states can be easily prepared by single-site addressing or using dissipation.
 It is worth to notice that the configuration with the positive charge at the top-right can be obtained from the one above by combining a $\pi$-rotation of the lattice, which puts the charges in the desired configuration but reverses the orientation, with the flip of the state of each link, $\ket + \leftrightarrow \ket -$, which restores the right orientation.
  Hence, the preparation of initial state to be driven in presence of charges does not imply much more effort than the preparation of the not-charged one.  

The other difference in preparation of the ground-state with charge with respect to case without is that the gauge condition to be enforced during the driving in order to minimize the error is different. Indeed, on the site where $\pm 1$ charge is placed the gauge condition reads   
\begin{equation}
A_{s_\pm} \ket \psi= e^{\pm i \alpha} \ket \psi. 
\end{equation}
This means that the dissipative procedure has to force the eigenvalue 1 of the operator $A^\prime_{s_\pm}= e^{\mp i \alpha} A_{s_\pm}$. 
By noticing that 
$$e^{\pm i\frac{\alpha}2\sigma_c^z} G(A)= e^{\pm i\frac{\alpha}2} G(A')= e^{\pm i\frac{\alpha}2}\left( \proj 0 \otimes \mathbb{1} + \proj 1 \otimes e^{\mp i \alpha} A\right),$$
we conclude that the correct gauge condition can be imposed via the modified version of dissipative sequence (\ref{UD})
\begin{equation}
U^\pm_D = \tilde G(A^\prime_{s_\pm})^{\dagger} U_{s_\pm}^{z,\varphi} \tilde G(A^\prime_{s_\pm}),\label{UDpm}
\end{equation}
where   $\tilde G(A^\prime_{s_\pm})= e^{i\frac {\pi}4 \sigma_c^y}e^{\pm i\frac{\alpha}2\sigma_c^z}G(A_{s_\pm})e^{-i \frac{\pi}4 \sigma_c^y}= e^{\pm i\frac{\alpha}2}\left( \mathbb{1}_c\otimes \frac {\mathbb{1}_{s_\pm}+ e^{\mp i \alpha} A_{s_\pm}}2 + \sigma_c^x\otimes \frac {e^{\mp i \alpha}A_{s_\pm} -\mathbb{1}_{s_\pm}}2 \right)$, as the phase $e^{\pm i\frac{\alpha}2}$ cancels out in the gate.

\subsection{Validation and usefulness of the simulator: absence of symmetry breaking and confinement}\label{veri}
Once the proposed experiment has been realized, one has to be able to  ensure that the simulator is indeed behaving   
as expected, and thus validate the results.
This is in general a non-trivial task \cite{Hauke11}. In the case of a gauge theory simulation, there are, however, at least two expected properties that one could use as a ``smoking gun'' of the desired physics.

First, one can check that the expectation value of any non-gauge invariant operator is zero as predicted by gauge invariance for well defined, i.e., non anomalous gauge theories like the gauge magnet under scrutiny.
The simplest of such operators are (twice) the Hermitian and antiHermitian part of $Z$, in the logical basis, $Z + Z^\dagger=\sigma^z$ and $-i(Z - Z^\dagger)=-\sigma^y$ , respectively. Verifying that the expectation value of these operators is zero for any link provides a simple necessary condition of the consistent realization of the gauge symmetry -- obviously, it is saying nothing about how good is the simulation of the dynamics.    
In order to make the check of gauge-invariance more compelling -- it is well known that while the expectation value of  local operators can be reasonably close to what craved, the state can still be orthogonal to the one desired-- 
we may consider similar operators on larger support such as non-intersecting string of links, $\bigotimes_l \sigma_l^{z(y)}$.  The vanishing of all the $\langle\bigotimes_l \sigma_l^{z(y)}\rangle$ would  provide  a stronger  confirmation about the gauge invariance of the prepared state.


It is worth to notice that, as these measurements are destructive, i.e., they are non-commuting with the gauge-condition and with the Hamiltonian, the (ground-)state of the model has to prepare again after any of these measurements.


The second important check is non destructive, and concerns the dynamics of gauge theories. As argued in section \ref{truu1}, a common mark to them is  charge confinement.
This can be tested by measuring the charge gap defined as the difference in energy between the ground-state with opposite $\pm 1$ charges and the vacuum. It requires the comparison of the ground-state energies in presence of $\pm 1$ static background charges at different distances. The vacuum may be thought as the limit configuration of zero displacement between the charges. As explained in  section \ref{truu1}, the different ground-state preparation amount in changing the gauge symmetry condition from (\ref{eq:gauge_cond}) to (\ref{eq:charges}) at the two sites where the charges are located. 
In presence of confinement, the dependence of the charge gap  upon the inter-charge distance (for the model proposed) has to be linear,  with the proportionality factor expressing the string tension, see (\ref{eq:string_tens}).
This measurement can be performed at both stages of the ground-state preparation. Indeed, after the dissipative stage, charges should already be confined and the string tension in known to be $\sigma=1/\sqrt{2}$. The value extracted from the experiment can, then, be used to check whether the first step of the state preparation has been successful. After the implementation of the ground-state of the full Hamiltonian, for instance, via the CRAB optimization,  charges should still be confined but with a different and unknown value of the string tension. At present,  this value cannot been determined with other means  for sufficiently large lattices (the string tension itself is not well-defined for small lattices). Hence,  it would be first interesting determination of the quantum simulation.

\subsection{Some remarks on the experimental set-up and summary of the experimental sequence}\label{summary}

For sake of clarity, we conclude this section by summarizing which are the key ingredients to build up the quantum simulator and which is the experimental schedule of the simulation.

\subsubsection{Requirements of the implementation}  

\begin{itemize}
\item
The first requirement for setting up the experiments is the choice of ultracold atoms suitable for the implementation of the mesoscopic Rydberg gate and for the load in the appropriate optical lattice. In the simplest scenario we may consider,  ensemble 
and control atoms are of the same atomic species, for instances Rubidium.  As a consequence, an ordinary square lattice, obtained by shining the atoms with two pairs of 
counterpropagating lasers in the X and Y direction, respectively, is sufficient to host all the atoms. If distinct atomic species for ensemble and control  are considered, a more complicated superlattice structure has to be employed, 
as, for separate, physical (link qubits represented by the ensemble atoms) and ancillary (control atoms) degrees of freedom occupy two square lattices at $45\,^{\circ}$ (see Fig. \ref{fig:simulator}).    
In any case, the lattice potential has to be sufficiently high that the free hopping is suppressed (compared with the time scale of the experiment): for simplicity, the atoms are assumed to be in Mott state with just one atom per site.  
\item
The second crucial requirement is single-site addressing, i.e., the capacity of (laser) manipulating the atom in each site individually. Single-site addressing is technically hard but possible \cite{Bakr09} in ultracold atom experiments. 
It is at the hearth of the functioning of mesoscopic Rydberg gate, see Fig. \ref{fig:ryd}. Furthermore, due to single-site addressing, the position in the lattice is sufficient by itself to distinguish control atoms from the ensemble ones.
\item
The third requirement is that the system be sufficiently cold such that the energy scale $E$ associated to the Hamiltonian can be resolved, i.e., $KT< E$ where $T$ is the temperature and $K$ the Boltzmann constant.  
The  energy scale  $E$, roughly speaking the normalization of the plaquette term (for convenient it has been fixed to 1 in our analysis),  is limited by the Trotter approximation to be much less than the inverse of the time $\Delta t$ in which one unitary step of time evolution is performed. Such time $\Delta t$ is determined by the number of sequential Rydberg gates $N_R$ needed to engineer the plaquette term evolution,  $\Delta t = N_R t_R$, as the further delay due to single-qubit rotations entering the process is negligible compared to the Rydberg gate delay $t_R$. Hence, it follows $N_R\ll t_R KT$. For state of the art experiments $t_R KT\gtrsim 10^2$, as $t_R$ is about few $\mu s$ and $T$ is about few tens of $nK$. Let us now compute $N_R$. The plaquette term decomposes in eight elementary monomials  of $\sigma^\mu$, $\mu=0,\dots,3$, $\sigma^0=\mathbb{1}$, as shown in  (\ref{pladeco}). Two of them, $(\sigma^z)^{\otimes 4}$ and $(\sigma^y)^{\otimes 4}$, are commuting when applied on contiguous plaquettes. The other six, $(\sigma^z\otimes\sigma^y)^{\otimes 2}$, $(\sigma^y\otimes\sigma^z)^{\otimes 2}$, $\sigma^z\otimes\sigma^y\otimes\sigma^y\otimes\sigma^z$, $\sigma^y\otimes\sigma^z\otimes\sigma^z\otimes\sigma^y$, $(\sigma^z)^{\otimes 2}\otimes(\sigma^y)^{\otimes 2}$, $(\sigma^y\otimes\sigma^z)^{\otimes 2}$ can be decomposed in twelve commuting monomials once applied on the alternated plaquettes. As the unitary evolution of each monomial involves two Rydberg gates, the total number of gates performed is $N_R= (2+ 6 \times 2)\times 2 = 28$, which is sufficiently small to satisfy the condition. Performing gate optimization, $N_R$ may be further reduced.      
\end{itemize}

 \subsubsection{Experimental sequence}
 
 \begin{enumerate}
 \item Loading of the ultracold atoms in the appropriate square (super)lattice as in Fig. \ref{fig:simulator}.
 \item Preparation of the initial appropriate product state  as in section \ref{sec:statepreparation}, by using dissipation via Rydberg gates (or dissipation combined with single-site manipulation of individual ensemble atoms). As discussed in section \ref{truu1}, the matter content of the final desired state to achieve (after the driving, see next point) determines which is the appropriate initial state to start with.  
 \item Adiabatic unitary evolution using Rydberg gates as explained in section \ref{sec:statepreparation}. Commuting operations can be performed on parallel. More efficient driving as the CRAB method can be adopted.
 \item Error minimization by enforcing the gauge condition during the driving via dissipation (see section \ref{diss}). The two procedures, driving and dissipation, can be applied for alternative intervals of time.
 \item Validation of the simulator and ``computation" of the simulation. Once the ground-state of the system (for the chosen matter content) has been achieved, it is characterized by measuring correlation functions. 
       Such operation can be performed by mapping the expectation value of the operator on the control atoms with the gate $\tilde G$, as defined in (\ref{readergate}). Indeed, the relevant operators to be evaluated are (linear combination of) monomials of $\sigma^\mu$, $Q=\sigma^{\mu_1}\otimes\dots\otimes\sigma^{\mu_k}$, both for the validation and the evaluation of the string tension. The latter requires the evaluation of the energy decomposed in the sum of commuting monomials with $k=4$ (as explained above). In practice, one starts with the control initialized in $\ket 0$. Once applied the unitary map $\tilde G$, the control will be in the linear superposition $\ket  0$ and $\ket 1$, reflecting which is the decomposition of the state in eigenvalues $\pm 1$ of $Q$ (see explanation below (\ref{readergate})). The control state can be measured in various ways, for instance using florescence \cite{Sherson10}. Afterwards, the original state of the simulator has to be prepared again, as the measurement alters it. In order to get enough statistics, the whole process has to be repeated several times. 
\end{enumerate}

\section{Conclusions and outlook}

In this work, we have introduced a new derivation of the  $U(1)$ gauge magnet in 2+1 dimension as a special case of a generic Abelian LTG, and discussed its phase diagram for a natural choice of the gauge invariant Hamiltonian.  The advantage of such derivation is that it shows explicitly that the U$(1)$ gauge magnet  may be viewed as a truncated formulation of the  standard Kogut-Susskind $U(1)$ LGT theory. In particular,  the amount of resources needed for the simulation is drastically reduced as the local Hilbert space shrinks from infinity to 2D (qubit). This allows to engineer a digital simulation of gauge magnet in optical lattices based on the latest developments in Rydberg gates.

In this context, we have proposed a protocol that allows for both  preparing the ground-state of the model and performing its out of equilibrium dynamics.
We also discuss two possible measurements that can provide {\it sine qua no} conditions to the correctness of simulation. In order to perform a successful simulation, one has   at least  check that the states obtained are gauge invariant and that static charges are confined. The determination of the string tension for the full-fledge Hamiltonian of the $U(1)$ gauge magnet would  also provide the first non-trivial result obtained from a quantum simulator in the area of gauge theories.

Our results pose interesting questions and open novel possibilities that deserve further studies. First of all, we (re)discover that  the simplest plaquette Hamiltonian gives rise to confined phase, as expected, but gapless. Such property distinguishes the gauge magnets from the ordinary U$(1)$ gauge theory in  2+1 (cf. \cite{polyakov_compact_1975}), and it is a consequence of novel gauge condition arising from the different choice of the Hilbert space (such choice, for instance, rules out the existence of short closed-string configurations).
How the choice of the group representation affects universal properties like phase diagrams and phase transitions is a fundamental question that raises naturally here and is asking for solutions.   

Similarly, our constructive approach to gauge magnets allows an infinite class of Hamiltonians built up of local terms to be considered. It is totally unknown, at the moment, whether novel phases and phase transitions are described by such models, and, for instance, whether the Polyakov phase could coexist in the same phase diagram with the gapless phase we have encountered in our study.      
The above questions are especially stimulating as nowadays novel tools in both classical simulation algorithms and in quantum simulation may be designed and employed to find a solution.

Furthermore,  another interesting line of investigation is the search for analogues of the gauge magnets formulations in 3+1 dimensions and/or for non-Abelian groups. In parallel, the introduction of the coupling to charged matter is very appealing. As relativistic fermions  are relatively easy to be simulated {\it analogically} in optical lattices, a combined digital/analogic simulation approach is desirable.

\acknowledgments{The idea of using quantum links for the quantum simulation of dynamical gauge theories originated  at KITP, from  discussions of ML with U.J.Wiese, whom we kindly thank. 
We are grateful to 
T. Caneva, I. Cirac, J. Rodriguez-Laguna, and G. Vidal for stimulating discussions.
We acknowledge the financial support from the
Spanish MEC project TOQATA (FIS2008-00784) and Consolider QOIT, Marie Curie project FP7-PEOPLE-2010-IIF
ENGAGES 273524, ERC Advanced Grant
QUAGATUA, EU STREP AQUTE, NAMEQUAM, and from the
Alexander von Humboldt Foundation.}

{\it Note added}. Little after the appearance of this work, \cite{Banerjee12} appeared
proposing a different way of simulating link model in optical lattices.

\begin{appendix}

\section{Gauge group $\mathbb{Z}_{2}.$}\label{sect:z2}

\subsection{The local Hilbert space}

The $\mathbb{Z}_{2}$ is the simplest LGT one could imagine (apart from the percolationg LGT of  \cite{gliozzi_random_2005}) where the orientation of the lattice is not important, since
there are only two different elements of the group, the first is $1$
and the other is $e$, which has the property that $e^{2}=ee^{-1}=1$.
The local Hilbert space with spin $1/2$ is the algebra of the group
and the representation matrices of the group coincide with the regular
representation of the group.
The regular representation is defined to be the representation in
which the matrices are constructed directly from the multiplication
table of the group. In order to get the representation of multiplication
by a given element, one substitute the given element in the multiplication
with one and all the others with zeros. For the $\mathbb{Z}_{2}$
case we have the following group multiplication table \begin{equation}
\begin{array}{c|cc}
 & 1 & e\\
\hline 1 & 1 & e\\
e^{-1}=e & e & 1\end{array}\end{equation}
 thus $X(1)=\mathbb{1}$, $X(e)=\sigma_{x}$. There is an important
theorem relating the regular representation with the irreducible representations.
This theorem states that the regular representation contains all the
irreducible representations a number of time equal to their dimensions \cite{tinkham_group_2003}.

In this case, we have two $A_{s}(g)$, 
$A_{s}(1)=\mathbb{1}$ and $A_{s}(e)=\sigma_{x}^{\otimes4}$. Thus the condition (\ref{eq:gauge_invariant_state}) is projecting out half of the $2^4$ states associated to a site.

\subsection{The gauge invariant Hilbert space}

In order to show that the condition (\ref{eq:gauge_invariant_state}) is neither trivial or empty, and how it can be implemented, we first consider the minimal lattice, made of only one plaquette. This is shown explicitly in
Fig. \ref{fig:Def} v). The gauge invariant Hilbert space is embedded
in the $2^{4}=16$ dimensional Hilbert space $\{\ket{l_{1}l_{2}l_{3}l_{4}}\}.$ Due to the periodic identification of of the links on the plaquette, 
 (\ref{eq:gauge_invariant_state}) applied for $s_{1}$ implies
\begin{equation}
\sigma_{x}^{l_{4}}\otimes\sigma_{x}^{l_{1}}\ket{l_{1}l_{4}}=\ket{l_{1}l_{4}}.
\end{equation}
Since $\sigma_{x}$ has eigenvalues $\pm1$. We have two possibilities
$\ket{l_{1}l_{4}}=\ket{++}$, $\ket{l_{1}l_{4}}=\ket{-- }$, where 
\begin{equation}
\sigma_{x}\ket +=\ket +,\ \sigma_{x}\ket -=-\ket -.
\end{equation}
By iteratively applying all the $A_{s}(g)$ on subsequent sites, one
realizes that there are two allowed states and thus a generic gauge
invariant state is a linear combination of them
\begin{equation}
\ket{\phi_{+}}=\ket{++++},\ \ket{\phi_{-}}=\ket{-- -- }.
\end{equation}
 This formalize the naive intuition (which has to be modified for finite systems with periodic boundary conditions \cite{tagliacozzo_entanglement_2011})  that gauge invariant states are related to the elementary plaquettes of the lattice.

\subsection{The gauge invariant operators}

The operators compatible with  gauge invariance constraints are
either
\begin{itemize}
\item product of $\sigma_{x}$ on arbitrary links of the lattice
\item product of $\sigma_{i\ne x}$ on closed paths
\end{itemize}
The simplest choice is then $\sigma_{z}^{l_{1}}\otimes\sigma_{z}^{l_{2}}\otimes\sigma_{z}^{l_{3}}\otimes\sigma_{z}^{l_{4}}.$
 Going back to the lattice formed by a single plaquette of Fig. \ref{fig:Def} v) we define the Hamiltonian 
\begin{equation}
H_0=-B_{P}=-\sigma_{z}^{l_{1}}\otimes\sigma_{z}^{l_{2}}\otimes\sigma_{z}^{l_{3}}\otimes\sigma_{z}^{l_{4}}, \label{eq:ham_deconf}
\end{equation}
that is the form that the generic Hamiltonian of  (\ref{eq:ham_ks}) on the chosen  lattice when $\theta = 0$.
In this simple example, we see that the ground-state is given by the
linear combination 
\begin{equation}
\ket{\psi_{0}^{B_{p}}}=\frac{1}{\sqrt{2}}\left(\ket{++++}+\ket{-- -- }\right),
\end{equation}
and the first excitation is given by
\begin{equation}
\ket{\psi_{1}^{B_{p}}}=\frac{1}{\sqrt{2}}\left(\ket{++++}-\ket{-- -- }\right).
\end{equation}

We also see that the two states are separated by an energy gap (that does not depend on the system size),
\begin{equation}
\Delta_2=2.
\end{equation}
On the other hand, if  we consider the opposit limit of the   Hamiltonian of  (\ref{eq:ham_ks}) and set $\theta = \pi / 2 $, we obtain
\begin{equation}
H_{\pi/2}=-\sum_{l_{i}}\sigma_{x}^{l_{i}}),\label{eq:ham_conf}
\end{equation}
 which ground-state becomes 
\begin{equation}
\ket{\psi_{0}^{\sigma_{x}}}=\ket{++++},
\end{equation}
and which  first excited state is 
\begin{equation}
\ket{\psi_{1}^{\sigma_{x}}}=\ket{-- -- }.
\end{equation}
Already in this simple example we can appreciate that in the two limit considered   of  (\ref{eq:ham_ks}) we are in a gaped phase but the ground-states 
of the Hamiltonian in  (\ref{eq:ham_conf}) and  (\ref{eq:ham_deconf}) have different entanglement properties. Indeed, the ground-state of the latter Hamiltonian is a product state,
while the one of the former is locally maximally entangled, since the reduced density matrix of a single link is proportional to the identity. This reasoning can be extended to larger systems since both Hamiltonians of (\ref{eq:ham_conf}) and (\ref{eq:ham_deconf}) are fixed point of the RG flow \cite{tagliacozzo_entanglement_2011}. The general Hamiltonian (\ref{eq:ham_ks}) is not in general a fixed point of the RG flow. From the above discussion it is easy to accept  the appearence of a phase transition at a given $\theta_c$ in between $\theta =0$ and $\theta = \pi /2 $ that separate the domain of attraction of the RG fixed point at $\theta =0$ usually called the topological-deconfined phase with the one of the RG fixed point at  $\theta = \pi/2$ usually called the confined phase.

\section{The case of the group $\mathbb{Z}_{3}$, prototype for the generic $\mathbb{Z}_{N}$}\label{sect:z3}

\begin{figure}[htb]
\includegraphics[width=10cm]{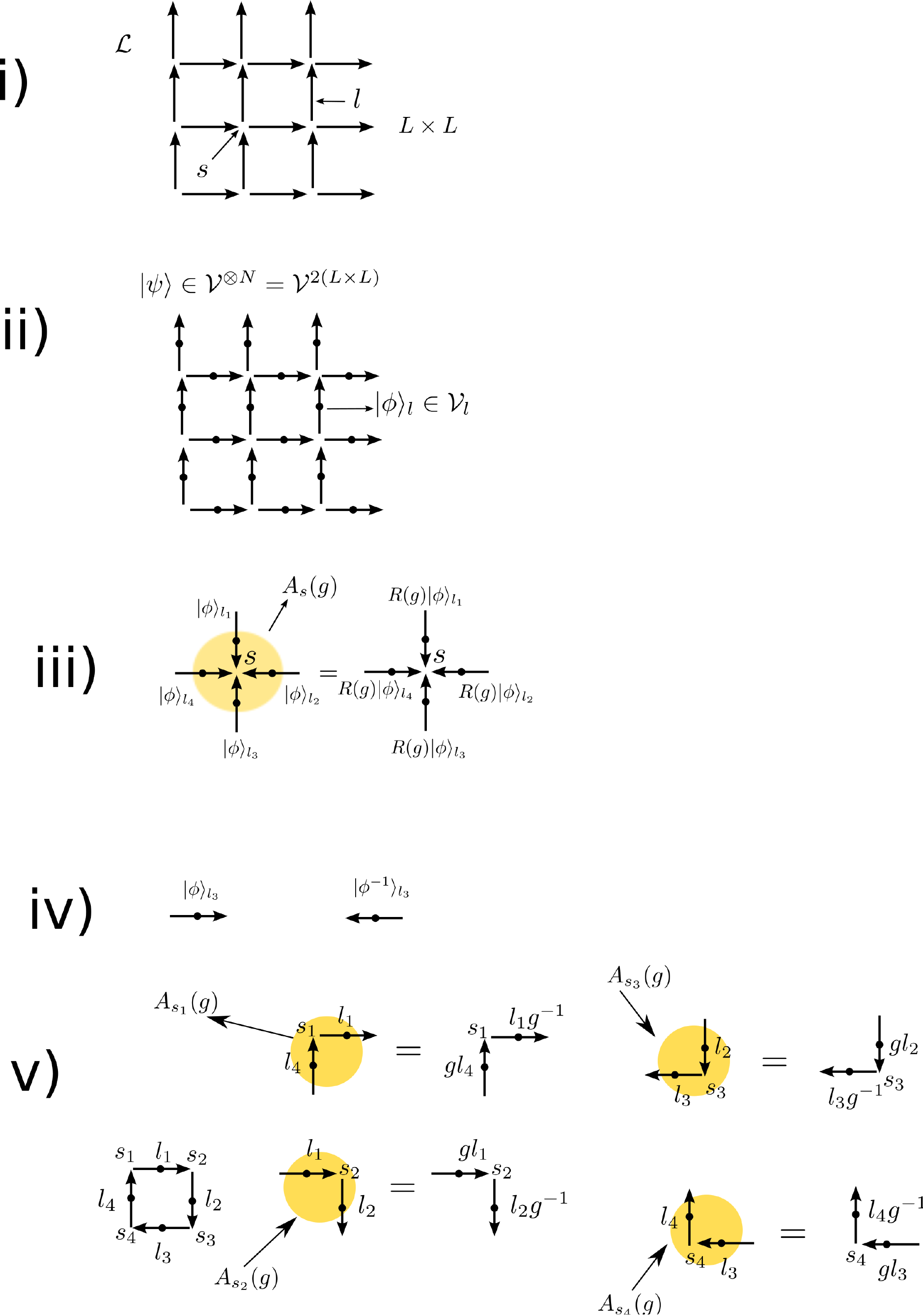}

\caption{i) Example of an oriented lattice $\mathcal{L}$ with $L$ sites $s$
and $N=2L$ links $l$. ii) A many-body state of the lattice is obtained
by defining a local Hilbert space $\mathcal{V}$ on each of the links.
A state of the lattice is then a state of the tensor product $\ket{\psi}\in{\cal H}_G=\mathcal{V}^{\otimes N}.$
iii) For each site $s$ of the lattice we can define operators $A_{s}(g)$
acting on the local Hilbert space ${{\cal H}_G}|_{s}=\mathcal{V}^{\otimes4}$, where $\mathcal{V}$ is the Hilbert space of a link ending in $s$. $A_{s}(g)$ multiplies each
link state $\ket{h_{l_{i}}}_s$ for the $X(g)$ or $X(g^{-1})$ depending on the fact
if the link is entering or exiting the particular site. Gauge invariance
is the requirement that the state $A_{s}(g)\bigotimes_i\ket{h_{l_{i}}}_s\rangle=\bigotimes_i\ket{h_{l_{i}}}_s\rangle\rangle$.iv)
A particular case is obtained when the local Hilbert space $\mathcal{V}$
is chosen to be the group algebra ${\cal C}(G)$. In this case, the
states are labeled by $\ket {g_{l_{i}}}_s$ and changing the orientation
of the link, corresponds to $\ket {h_{l_{i}}}_s\to\ket{{h^{-1}}_{l_{i}}}_s.$
v) A simple example of a one plaquette lattice, discussed in the main text, is worked out in details.
It consists of four sites, $s_{1}\cdots s_{4}$
and four links $l_{1}\cdots l_{4}$. The four elementary gauge transformations
related to the four sites are explicitly introduced. \label{fig:Def}}

\end{figure}

The next step of complication is obtained by considering a $\mathbb{Z}_{3}$.
The main difference with respect to the $\mathbb{Z}_2$ theory is that the group algebra of $\mathbb{Z}_3$ has dimension 3 - $N$ for a generic $\mathbb{Z}_N$ - and not all the elements of the group are  the inverse of themselves. Hence, the orientation of the lattice plays a central role.  From now on, we will consider two different basis,
the group algebra basis 
$\ket g $,
 and the basis in which the regular representation becomes the direct
sum of the irreducible representations $ \ket t$.

In the $\ket g$-basis the $X(g)$s read
\begin{equation}
\begin{array}{c|ccc}
 & 1 & e_{2}=e_{1}^{-1} & e_{1}=e_{2}^{-1}\\
\hline 1 & 1 & e_{2} & e_{1}\\
e_{1} & e_{1} & 1 & e_{2}\\
e_{2} & e_{2} & e_{1} & 1\end{array},
\end{equation}
\begin{align}
X(1)&=\mathbb{1}\,,\label{eq:R1}
\\
X(e_{1})&=\left(\begin{array}{ccc}
0 & 0 & 1\\
1 & 0 & 0\\
0 & 1 & 0\end{array}\right),\label{eq:R2}
\\
X(e_{2})&=\left(\begin{array}{ccc}
0 & 1 & 0\\
0 & 0 & 1\\
1 & 0 & 0\end{array}\right).\label{eq:R3}
\end{align}
As done previously for $\sigma_x$, we can diagonalize all the $X$ and we pass to the basis $
\proj{t_{i}}$, $i=1\dots 3$. In this basis
\begin{align}
X(1)&=diag\{1,1,1\},\\
X(e_{1})&=diag\{1,e^{\frac{4\pi}{3}i},e^{\frac{2\pi}{3}i}\},\\
X(e_{2})&=diag\{1,e^{\frac{2\pi}{3}i},e^{\frac{4\pi}{3}i}\}.
\end{align}

We then can find the gauge invariant states in the $\ket t$-basis
\begin{equation}
\ket{t_{1,}t_{1},t_{1},t_{1}},\ket{t_{2},t_{2},t_{2},t_{2}},\ket{t_{3},t_{3},t_{3},t_{3}}.\end{equation}

Now, let us focus on the construction of the plaquette of the Hamiltonian, taking properly in account the orientation of the links. In order to do so, we have to specify the form of the $Z$ operator for the group $\mathbb{Z}_{3}$, which generalizes the $\sigma_{z}$ employed for $\mathbb{Z}_{2}$. In
the $\ket g$-basis, it is \begin{equation}
Z=\left(\begin{array}{ccc}
1 & 0 & 0\\
0 & e^{\frac{2\pi i}{3}} & 0\\
0 & 0 & e^{\frac{4\pi}{3}i}\end{array}\right).\label{eq:Z}\end{equation}

It has the following properties
\begin{equation}
ZX(e_{1})=e^{\frac{2\pi}{3}i}X(e_{1})Z,\,\,\,\,\,\, ZX(e_{2})=e^{\frac{4\pi}{3}i}X(e_{2})Z.\label{eq:comm_rel}
\end{equation}

If we orient the plaquettes in such a way that they always have
one entering and one exiting link for each site (either anti-clockwise
or clockwise) the plaquette will meet a gauge transformation
on two consecutive links but the orientation is such that the gauge
transformation on these links will look like $X(e)\otimes X(e^{-1})$
. Since the orientation of the plaquette is different
from the orientation of the gauge transformation given in Fig. \ref{fig:Def}
iii) we are multiplying one of the two links for $X(e^{-1})$ instead
than $X(e)$. Hence, the only possibility to obtain an operator commuting
with all the gauge transformations is to take again a product of $Z$ as
\begin{equation}
[Z\otimes Z,X(e)\otimes X(e^{-1})]=0,
\end{equation}
accordingly to (\ref{eq:comm_rel}). It follows that the correspondent Hamiltonian term, written on the anti-clockwise or clockwise oriented
plaquettes (as the one of Fig. \ref{fig:Def} v) ), reads 
\begin{equation}
H_P=-\frac 12 \sum_p \left(Z_{l_{1}}Z_{l_{2}}Z_{l_{3}}Z_{l_{4}} + Z_{l_{1}}^{\dagger}Z_{l_{2}}^{\dagger}Z_{l_{3}}^{\dagger}Z_{l_{4}}^{\dagger}\right).
\end{equation}
 This is, as expected,  the  explicit realization of  Hamiltonian of  (\ref{eq:ham_ks}) with $\theta=0$ for the group $\mathbb{Z}_{3}$. It is important to notice that now the $l_{1}\cdots l_{4}$ are the
links around a given plaquette in the order of Fig. \ref{fig:Def}
v). Once we move back to the standard orientation of the 2D space of
Fig. \ref{fig:Def} ii) we recover the standard form \cite{kitaev_fault-tolerant_1997}
\begin{equation}
H_P=-\frac 12 \sum_p \left( Z_{l_{1}}^{\dagger}Z_{l_{2}}^{\dagger}Z_{l_{3}}Z_{l_{4}}+Z_{l_{1}}Z_{l_{2}}Z^{\dagger}_{l_{3}}Z^{\dagger}_{l_{4}}\right).
\end{equation}

Once the basis is rotated from $\ket g$ to $\ket t$, the form of the $Z$ operator is determined by (\ref{eq:comm_rel}) to be
\begin{equation}
Z:\ket{t_{i}}\to\ket{t_{i+1}},\end{equation}
and \begin{equation}
Z^{\dagger}: \ket{t_{i}}\to\ket{t_{i}-1},\end{equation}
i.e., $X(e_1)\to Z$, and $Z\to X(e_1)$. 

Again we can consider the ground-state of
the plaquette operator for a lattice made of a  single plaquette as in Fig. \ref{fig:Def}
v). It turns out that 
\begin{gather}
Z_{l_{1}}Z_{l_{2}}Z_{l_{3}}Z_{l_{4}}: \ket{t_{i}t_{i}t_{i}t_{i}}\to\ket{t_{i+1}t_{i+1}t_{i+1}t_{i+1}},\\
Z_{l_{1}}^{\dagger}Z_{l_{2}}^{\dagger}Z_{l_{3}}^{\dagger}Z_{l_{4}}^{\dagger}:\ket{t_{i}t_{i}t_{i}t_{i}}\to\ket{t_{i-1}t_{i-1}t_{i-1}t_{i-1}}.
\end{gather}
It is easy to realize that the eigenvalues of $H_P$ are $-1$ or $+1/2$ and
its ground-state on a single plaquette is given
by 
\begin{equation}
\ket{\psi_{0}}=\frac 1{\sqrt 3}\sum_{i}\ket{t_{i}t_{i}t_{i}t_{i}}.
\end{equation}
Here, again we discover the emergence of gauge invariant plaquette states. Their number is now $3$ and clearly coincide with the $|G|$, the cardinality of the group. The plaquette  Hamiltonian is still gapped, but the gap, $\Delta_3=3/2$ is now smaller than in the $\mathbb{Z}_{2}$ case. We can again study the opposite limit of   (\ref{eq:ham_ks})  with $\theta=\pi /2$. There we see that all spins are aligned in the $t_1$ direction. As before, by passing to larger lattices, we can repeat the same reasoning and we find two different phases. One in  which the ground-state in a product state and the other in which it is robustly entangled. It is also known that by increasing the rank of the group a third phase appear. Here, we will not study that phase and refer the interested reader to the literature \cite{horn_hamiltonian_1979}.
\end{appendix}

\end{document}